\immediate\write18{makeindex \jobname.nlo -s nomencl.ist -o \jobname.nls}
 \documentclass[final,5p,times,twocolumn]{elsarticle}

\usepackage{url}
\usepackage{caption}
\usepackage{subcaption}
\usepackage{diagbox}

\usepackage{array}
\usepackage{multirow}
\newcolumntype{P}[1]{>{\centering\arraybackslash}p{#1}}
\usepackage{xtab,afterpage,booktabs}
\usepackage{xtab,afterpage}
\usepackage{amssymb}
\usepackage{amsmath, nccmath}
\usepackage{geometry}
\usepackage{physics}
 \usepackage{ltablex}
\usepackage[flushleft]{threeparttable}
\usepackage{tikz}
\usetikzlibrary{positioning, fit, arrows.meta, shapes}
\usepackage{subcaption}
\usepackage{stackengine}
\usepackage{mwe}
\usepackage{float}
\usepackage{nomencl}
\usepackage{amsmath}
\usepackage{tabularx}
\usepackage{adjustbox}
\usepackage{siunitx}
\usepackage{booktabs}
\usepackage{blindtext}
\usepackage{hyperref}
\usepackage[labelsep=space]{caption}  
\usepackage{enumitem}
\newlist{steps}{enumerate}{1}
\setlist[steps, 1]{resume,leftmargin=*,label = \textbf{Step \arabic*}:}
\usepackage{rotating}
\usepackage{array}
\usepackage{multirow}

\usepackage{graphicx}
\usepackage[linesnumbered,ruled,vlined]{algorithm2e}

\SetCommentSty{mycommfont}

\SetKwInput{-KwInput}{Input}                
\SetKwInput{KwOutput}{Output}              

\newlength\mylen

\usepackage{amssymb}
\usepackage{pifont}
\newcommand{\cmark}{\ding{51}}%
\newcommand{\xmark}{\ding{55}}%
\journal{Energy}
\usepackage{multicol}
\renewcommand*\nompreamble{\begin{multicols}{2}}
\renewcommand*\nompostamble{\end{multicols}}
\usepackage{etoolbox}
\renewcommand\nomgroup[1]{%
  \item[\bfseries
  \ifstrequal{#1}{P}{}{%
  \ifstrequal{#1}{Z}{Abbreviations}{}}%
]}
\usepackage[linewidth=0.5pt]{mdframed}
\newlength{\nomitemorigsep}
\makenomenclature

\usepackage{etoolbox}

\begin{document}


\makeatletter
\def\ps@pprintTitle{%
\def\@oddfoot{\footnotesize\itshape Accepted for publication in \@journal\hfil}}
\makeatother

\begin{frontmatter}



 \title{Day-ahead electricity price prediction applying hybrid models of LSTM-based deep learning methods and feature selection algorithms under consideration of market coupling}
 \author{Wei Li\corref{cor1}\fnref{mymainaddress}}
 \ead{wei.n.li@ntnu.no}
\author{Denis Becker\fnref{mymainaddress}}
 \cortext[cor1]{Corresponding author}
 \address[mymainaddress]{NTNU Business School, Norwegian University of Science and Technology, 7491 Trondheim, Norway\fnref{label3}\vspace{-3em}}




\begin{abstract}
The availability of accurate day-ahead electricity price forecasts is pivotal for electricity market participants. In the context of trade liberalisation and market harmonisation in the European markets, accurate price forecasting becomes difficult for electricity market participants to obtain because electricity forecasting requires the consideration of features from ever-growing coupling markets. This study provides a method of exploring the influence of market coupling on electricity price prediction. We apply state-of-the-art long short-term memory (LSTM) deep neural networks combined with feature selection algorithms for electricity price prediction under the consideration of market coupling. LSTM models have a good performance in handling nonlinear and complex problems and processing time series data. In our empirical study of the Nordic market, the proposed models obtain considerably accurate results. The results show that feature selection is essential to achieving accurate prediction, and features from integrated markets have an impact on prediction. The feature importance analysis implies that the German market has a salient role in the price generation of Nord Pool. 
\end{abstract}



\begin{keyword}
Deep learning  \sep Electricity price forecasting (EPF)  \sep Electricity market coupling \sep Feature selection \sep Long short-term memory (LSTM) \sep The Nord Pool system price


\end{keyword}

\end{frontmatter}


\section{Introduction}\label{S:1}
Over the last two decades, worldwide energy markets have experienced a transition towards deregulation and harmonisation \citep{Weron2006}. Under trade liberalisation, the traditional vertically integrated power utilities are replaced with decentralised business entities whose targets are to maximise their profits. Consequently, a growing number of market participants are exposed to intense competition, and their need for suitable decision support models to increase margins and reduce risk has significantly increased \citep{Bunn2004}. Thus, the availability of accurate day-ahead electricity price forecasts is vital for market participants to adjust production plans and to perform effective bidding strategies to make an economic profit. However, due to the productive structure and characteristics of electricity prices, highly accurate forecasting is quite challenging \citep{Nogales2002, Bunn2000}. With the increasing integration of electricity markets, making accurate forecasts becomes even more difficult in the complex and integrated system. This is because the forecasting of electricity prices needs to consider a large number of factors from an ever-growing number of interconnected, neighbouring power markets. These factors include electricity prices, production, consumption, and other important features that influence cross-border electricity markets.

\begin{table*}[t]
\begin{mdframed}
\printnomenclature
\end{mdframed}
\end{table*}
 Numerous research efforts have contributed to the exploitation and development of advanced technologies for day-ahead electricity price forecasting (EPF), aimed at highly accurate forecasting results \citep{WERON20141030, NOWOTARSKI20181548}. A considerable amount of literature has been devoted to EPF models, which can be classified into the following five categories \citep{WERON20141030}: multi-agent \cite{VENTOSA2005897, KIOSE20152731}, fundamental \citep{Burger2007,  WERON20141030}, reduced-form \citep{ISLYAEV2015144, WERON2008744}, statistical \citep{CONEJO2005435, Misiorek2006103, Gonzalez2018}, and computational intelligence (CI) models \cite{CATALAO20071297, KELES2016218,Peter20172277}. Compared with the other four traditional models, CI models are regarded as state-of-the-art techniques, and their superior performance contributes to their prevalence in EPF in recent years. In particular, deep neural networks (DNNs) have gradually become the most avant-garde CI approach in other disciplines \cite{Geoffrey2017, Bahdanau2014, LI2018340} and entered the scientific research field related to EPF.
 
DNNs are often categorised into three main classes: feed-forward neural networks (FNNs), recurrent neural networks (RNNs), and convolutional neural networks (CNNs). Different types of DNNs are used to solve different problems. For time series prediction, RNNs have achieved superior performance by building extra mappings to hold relevant information from past inputs. The long short-term memory (LSTM) and gated recurrent units (GRUs) are important variants of this kind of network, which overcome the vanishing gradient problem of RNNs \cite{Bengio1994157}. Compared with GRUs, LSTM  is more accurate on the dataset using long sequences. Due to the superiority of LSTM in time series forecasting, researchers have gradually paid attention to its application in EPF \citep{LAGO2018386, CHANG2019115804, Kuo2018}. However, as with other DNNs, when LSTM models are applied to high-dimensional data, a critical issue occurs, known as the curse of dimensionality \cite{hastie2005}. This means that, with a large amount of features,\footnote{In machine learning, features are individual independent variables as input in a model.} the performance of LSTM will degrade because of overfitting \cite{Li201745}. Thus, LSTM cannot be employed directly for electricity price prediction with a large number of features as input under consideration of market coupling. While some researchers have attempted to involve explanatory variables from integrated markets to make a prediction of electricity price by neural networks \cite{ZIEL2015430, PANAPAKIDIS2016132, LAGO2018890}, no existing research has investigated the state-of-the-art LSTM-based deep neural networks for this purpose. Besides, some research starts to pay attention to the influence of the market integration on Nord Pool \cite{URIBE2020118368, Marcjasz2020, Johannesen2019}. However, efficient ways to utilise the ever-growing information from the electricity market integration for the Nordic EPF have yet to be explored.

Typically, feature selection is an efficient way to avoid the curse of dimensionality. It is the process of selecting a subset of relevant attributes in the dataset when developing a predictive model. It can reduce the computation time, improve model prediction performance, and help to get a better understanding of the dataset \cite{CHANDRASHEKAR201416}. The ideal feature selection is to search the space of all variable subsets with an algorithm, which is impractical except for quite small sized feature spaces. However, as the space of variables subset grows exponentially with the number of variables, heuristic search methods are commonly used to search for an optimal subset \cite{Sanz2018}. The current research on feature selection algorithms can be categorised as filter, wrapper, and embedded methods \cite{CHANDRASHEKAR201416}. In particular, the filter methods use a proxy measure to estimate a feature subset before training a prediction model. Pearson’s Correlation (PC) is a typical indirect assessing measure for the regression problem \cite{Guyon20031532}. In contrast, the wrapper methods evaluate selected feature subsets by employing a predictive model directly. Each subset is used to train a new forecasting model, and the optimisation method is used to search for the best performing model in the process of feature selection. The embedded methods can implement an automatic feature selection in the process of estimating the parameters of predictive models. This means this catch-all group of techniques performs the process of feature selection during the training of the model. The particle swarm optimisation combined with the extreme learning machine method (PSO-ELM) and genetic algorithm combined with the extreme learning machine method (GA-ELM) are two typical wrapper-based methods. They have been widely used for various feature selection problems \cite{CHEN20061685, Nguyen20163927, Shang20163821, ZHOU2020117894, Krishnan2019525, Luo20181}. Guyon et al. \cite{Guyon2002} proposed another popular wrapper approach, known as recursive feature elimination combined with support vector machine for regression (RFE-SVR). The Lasso regression method is one of the most popular embedded feature selection methods proposed by Tibshirani \cite{Tibshirani1996267}.
 
\subsection{Contributions}
To the best of our knowledge, no existing study considers how to apply LSTM models in an integrated market EPF and detect the impact of the features from cross-border markets on EPF. To fill this scientific gap, we propose three hybrid architectures of LSTM-based deep learning predictive models combined with advanced feature selection algorithms: the two-step hybrid architecture, the autoencoder hybrid architecture, and the two-stage hybrid architecture. Different feature selection methods have different selection mechanisms, which will lead to different sets of selected features. To explore the influence of different feature selections on LSTM-based EPF, we employed five feature selection algorithms,  PC, PSO-ELM, GA-ELM, RFE-SVR, and the Lasso regression method, in the case study of Nord Pool and its neighbouring, interconnected countries. The main contributions of this study are as follows:

\begin{enumerate}
\item We compare and analyse the forecasting performance of the proposed models in the case study of the Nord Pool system price forecasting, considering six integrated markets (sixty-two features). The results indicate that the cross-border markets influence the Nordic electricity price formation. As the rapid market coupling development in Europe, we show that it is necessary to consider cross-border information for EPF in future studies.
\item We introduce three architectures of hybrid LSTM-based deep neural networks for EPF and conclude that different feature selection algorithms yield divergent subsets of features, which, in turn, affect the prediction accuracy of the proposed LSTM models. In addition, the results show that hybrid models are an efficient way to deal with the ever-growing information and obtain accurate prediction results in cross-border markets.
\item We employ a game theoretical approach (SHapley Additive exPlanations) to explore the relevance of various cross-border features in EPF. The analysis of Shapley values increases the transparency of the prediction and provides advice for policy makers and market participants.
\end{enumerate}

The remainder of this paper is organised as follows. Section 2 describes the dataset used in this research. In Section 3, we present the methodology. Section 4 describes the model training and introduces evaluation criteria applied in the empirical study. Section 5 reports the forecasting results of the implemented models. Finally, Section 6 concludes the paper and proposes future research developments.
 
\section{Data description} \label{S:3}
The Nordic system price is the central reference price in the Nordic electricity market. It is used as a settlement price for the derivatives market. Each hourly system price is calculated by Nord Pool based on all bids and offers posted in Nordic bidding zones, which is referred to as a market-clearing price, without taking into account any congestion restrictions. The daily system price represents the arithmetic average of the 24 hourly prices. This paper discusses and evaluates several hybrid LSTM-based  approaches for the prediction of the Nordic hourly and daily system prices.

Previous empirical research on the prediction of electricity prices has considered information from both price and supply/demand sides. To find out what matters when predicting the day-ahead Nordic system price in coupling markets, we also included the electricity exchange between Nord Pool and its integrated countries. The Russian electricity market is excluded because it differs significantly from European models. To consider the influence of the correspondence between electricity flow and capacity, we introduced a new daily feature, namely the cross-border flow deviation. It can be calculated as $\sigma_{FD} = \sqrt{{\sum_{i=1}^N (X_i-\mu_i)^2}/{N}}$, where $X_i$ is the hourly electricity flow, $\mu_i$ is the hourly expected exchange capacity, and $N$ stands for 24 hours. 

In summary, we consider eight categories of input features: day-ahead price, production, production prognosis, consumption, consumption prognosis, currency exchange rate, cross-border electricity flow, and flow deviation. The first five are the basic features from local markets for predicting electricity price. Some historical/predictive information, such as weather and human social activities, does not directly impact electricity price but influences the supply/demand for electricity, incorporated in those five fundamental variables. The last three are the features spawned from cross-border trade.

\nomenclature[P]{\(\sigma_{FD}\)}{Cross-border flow deviation}
\nomenclature[P]{\(\mu\)}{Expected electricity exchange capacity}
\nomenclature[P]{\(X, Y\)}{Variable}

\subsection{Data}
\begin{figure}[t!]
\centering
\includegraphics[height=6cm]{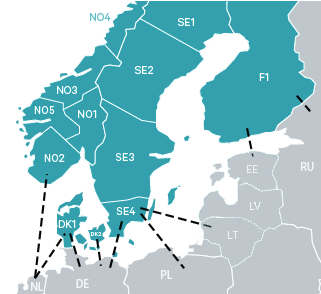}
 \caption{\centering Overview of the Nord Pool market coupling.}\label{f1}
\end{figure}

We collected data from the Nord Pool,\footnote{Nord Pool: https://www.nordpoolgroup.com/} Thomson Reuters Eikon,\footnote{Thomson Reuters Eikon: https://eikon.thomsonreuters.com/} and Entsoe.\footnote{Entsoe: https://transparency.entsoe.eu/} The available time series ranges from 01/01/2015 to 31/12/2019. Nord Pool provides cross-border transmissions with Germany (DE), the Netherlands (NL), Lithuania (LT), Estonia (EE), Poland (PL), and Russia (RU). The map in Figure \ref{f1} shows both the Nord Pool markets as well as transmissions (black dashed lines) between the Nord Pool and its coupling bidding areas. There are five bidding zones in Norway (NO1, NO2, NO3, NO4, and NO5), four in Sweden (SE1, SE2, SE3, and SE4), and two in Denmark (DK1 and DK2), and one in Finland (FI). Since the transmissions between DK1 and NL started at 01/09/2019, the data series is not sufficient for the application of deep learning models. Besides, the electricity exchange between SE4 and LT started at 09/12/2015. Therefore, the entire available dataset employed in this study ranges from 09/12/2015 to 31/12/2019. The features included in the dataset are shown in Table \ref{t1}. The hourly data is converted into the daily data by the arithmetic average (e.g., price) or the aggregate (e.g., flow). 

\subsection{Cross-border electricity transmission}
\begin{centering}
\begin{figure*}[h!]
\centering
\includegraphics[width =18cm, height = 9.5cm]{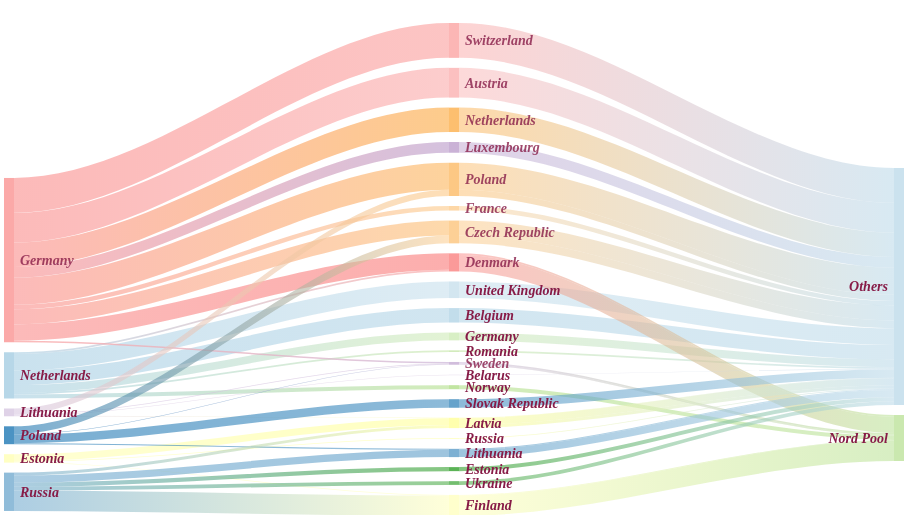}
\caption{\centering The electricity cross-border transmission from the coupling countries to Nord Pool.}\label{f_flow}
\end{figure*}
\end{centering}

\begin{figure}[h!]
\centering
\includegraphics[width =5cm]{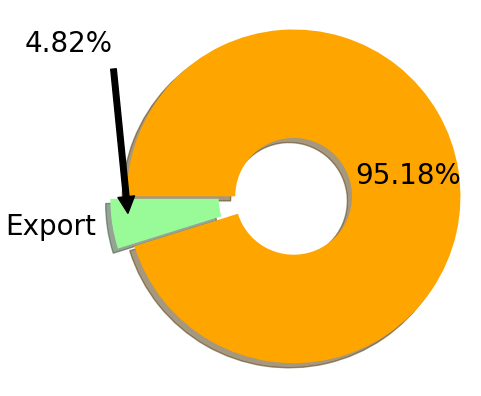}
\caption{The percentage of the Nord Pool production for exporting.}\label{f_e}
\end{figure}

Figure \ref{f_flow} shows the electricity exports from Germany, the Netherlands, Lithuania, Poland, and Russia in 2019.\footnote{Fraunhofer ISE provides the electricity exchange data of Germany/Europe: https://www.energy-charts.de/} The exports to the Nord Pool comprised 16.03\% of the whole exports from these coupling countries. In Figure \ref{f_e}, we can see that the electricity exports of the Nord Pool comprised 4.82\% of its total production in 2019. The EU aims to achieve 15\% interconnection capacity in 2030 for each EU country \cite{Greenfish2019}. 

\xentrystretch{-0.26}
\begin{center}
\topcaption{The features included in the dataset.\label{t1}}
\tablefirsthead{\toprule Feature & Description (Units) & Data Source \\ }
\tablehead{%
\multicolumn{3}{c}%
{{\bfseries  Continued from previous column}} \\
\toprule
Feature & Description (Units) & Data Source\\ \midrule}
\tabletail{%
\midrule \multicolumn{3}{r}{{Continued on next column}} \\ \midrule}
\footnotesize{
\begin{xtabular*}{.5\textwidth}{ll@{\extracolsep{\fill}}l}
\hline
 F1 & System Day-ahead price 1-Lag (EUR/MWh) & Nord Pool \\ 
 F2 & SE1 Day-ahead price (EUR/MWh) & Nord Pool \\ 
 F3&SE2 Day-ahead price (EUR/MWh)  & Nord Pool\\
 F4&SE3 Day-ahead price (EUR/MWh) & Nord Pool\\
 F5&SE4 Day-ahead price (EUR/MWh)  & Nord Pool\\
 F6&FI Day-ahead price (EUR/MWh) & Nord Pool\\
 F7&DK1 Day-ahead price (EUR/MWh)  & Nord Pool\\
 F8&DK2 Day-ahead price (EUR/MWh) & Nord Pool\\
 F9&NO1 Day-ahead price (EUR/MWh) & Nord Pool\\
 F10&NO2 Day-ahead price (EUR/MWh) & Nord Pool\\
 F11&NO3 Day-ahead price (EUR/MWh) & Nord Pool\\
 F12&NO4 Day-ahead price (EUR/MWh) & Nord Pool\\
 F13&NO5 Day-ahead price (EUR/MWh) & Nord Pool\\
 F14&EE Day-ahead price (EUR/MWh) & Nord Pool\\
 F15&LT Day-ahead price (EUR/MWh) & Nord Pool\\
 F16&PL Day-ahead price (PLN/MWh) & Thomson Reuters Eikon\\
 F17&DE Day-ahead price (EUR/MWh)  & Thomson Reuters Eikon\\
 F18&NL Day-ahead price (EUR/MWh)  & Thomson Reuters Eikon\\
 F19&Nordic production (MWh)& Nord Pool\\
 F20&EE production (MWh)& Nord Pool\\
 F21&LT production (MWh)& Nord Pool\\
 F22&PL production (MWh)& Entsoe\\
 F23&DE production (MWh)& Entsoe\\
 F24&NL production (MWh)& Entsoe\\
 F25&Nordic production prognosis (MWh)& Nord Pool\\
 F26&EE production prognosis (MWh)& Nord Pool\\
 F27&LT production prognosis (MWh)& Nord Pool\\
 F28&PL production prognosis (MWh)& Entsoe\\
 F29&DE production prognosis (MWh)& Entsoe\\
 F30&NL production prognosis (MWh)& Entsoe\\
 F31&Nordic consumption (MWh)& Nord Pool\\
 F32&EE consumption (MWh)& Nord Pool\\
 F33&LT consumption (MWh)& Nord Pool\\
 F34&PL consumption (MWh)& Entsoe\\
 F35&DE consumption (MWh)& Entsoe\\
 F36&NL consumption (MWh)& Entsoe\\
 F37&Nordic consumption prognosis (MWh)& Nord Pool\\
 F38&EE consumption prognosis (MWh)& Nord Pool\\
 F39&LT consumption prognosis (MWh)& Nord Pool\\
 F40&PL consumption prognosis (MWh)& Entsoe\\
 F41&DE consumption prognosis (MWh)& Entsoe\\
 F42&NL consumption prognosis (MWh)& Entsoe\\
 F43&EUR/NOK & Nord Pool\\
 F44&EUR/SEK & Nord Pool\\
 F45&EUR/DKK & Nord Pool\\
 F46&EUR/PLN & Thomson Reuters Eikon\\
 F47&NO2 $\leftrightarrow$ NL flow (MWh)& Nord Pool\\
 F48&DK1 $\leftrightarrow$ DE flow (MWh)&  Nord Pool\\
 F49&DK2 $\leftrightarrow$ DE flow (MWh)&  Nord Pool\\
 F50&SE4 $\leftrightarrow$ DE flow (MWh)&  Nord Pool\\
 F51&SE4 $\leftrightarrow$ PL flow (MWh)&  Nord Pool\\
 F52&SE4 $\leftrightarrow$ LT flow (MWh)&  Nord Pool\\
 F53&FI $\leftrightarrow$ EE flow (MWh)&  Nord Pool\\
 F54&FI $\leftrightarrow$ Russia flow (MWh)&  Nord Pool\\
 
 F55&NO2 $\leftrightarrow$ NL flow deviation& Calculation\\
 F56&DK1 $\leftrightarrow$ DE flow deviation& Calculation\\
 F57&DK2 $\leftrightarrow$ DE flow deviation& Calculation\\
 F58&SE4 $\leftrightarrow$ DE flow deviation& Calculation\\
 F59&SE4 $\leftrightarrow$ PL flow deviation& Calculation\\
 F60&SE4 $\leftrightarrow$ LT flow deviation& Calculation\\
 F61&FI $\leftrightarrow$ EE flow deviation& Calculation\\
 F62&FI $\leftrightarrow$ Russia flow deviation& Calculation\\
\bottomrule
\end{xtabular*}}
\end{center}
\normalsize

\section{Methodology}
\subsection{LSTM}
\begin{figure}[h!]
  \centering
\includegraphics[, height=6cm]{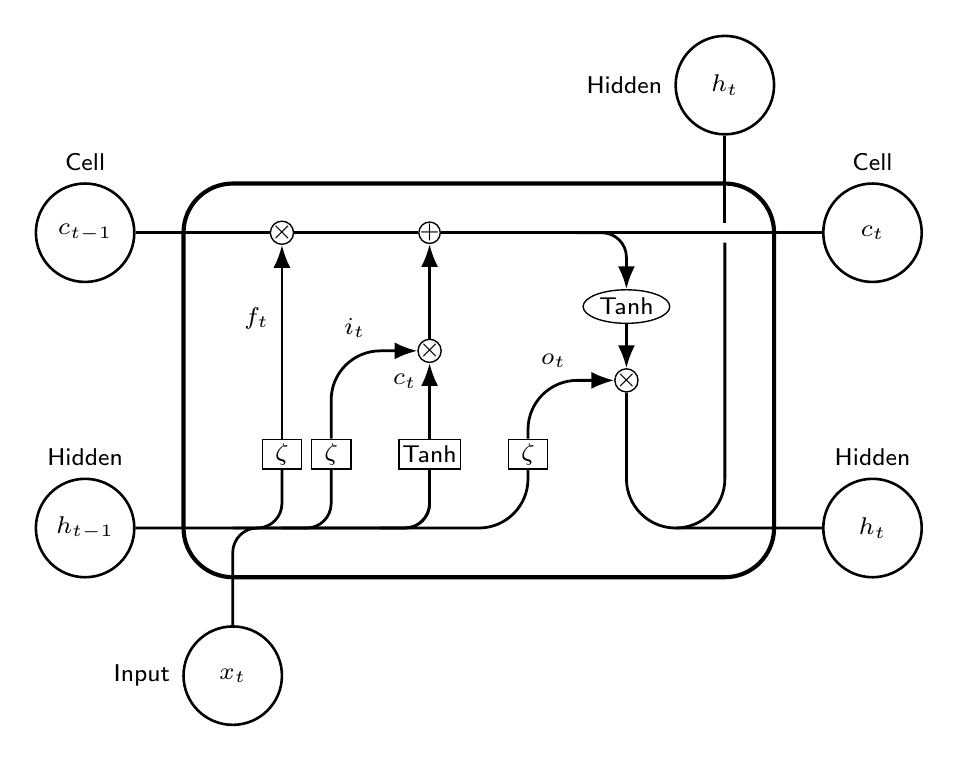}
 \caption{\centering LSTM cell.}\label{f7}
\end{figure}

The LSTM architecture was initially introduced by \cite{Hochreiter1997} and has since been enhanced by other researchers to achieve better performance  \cite{Gers2000, GRAVES2005602, Kyunghyun2014}. An LSTM network is a special kind of recurrent neural network that is capable of learning long-term dependencies. Unlike simple RNNs, an LSTM network has built-in mechanisms that control how information is memorised or abandoned throughout time. The architecture of the LSTM network is shown in Figure \ref{f7} and is defined by the following system of equations \cite{Graves2013}:

\begin{flalign}
&f_t = \zeta(W_{xf} x_t+W_{hf} h_{t-1} +W_{cf} c_{t-1} + b_f) \\
&i_t = \zeta(W_{xi} x_t+W_{hi} h_{t-1} +W_{ci} c_{t-1} + b_i) \\
&o_t = \zeta(W_{xo} x_t+W_{ho} h_{t-1} +W_{co} c_{t-1} + b_o) \\
&c_t = f_t \otimes c_{t-1} + i_t\otimes  \text{Tanh}(W_{xc} x_t+W_{hc} h_{t-1} + b_c) \\
&h_t = o_t \otimes  (c_t)
\end{flalign}
\noindent where $f_t$, $ i_t$, $o_t$, $c_t$, and $h_t$ indicate the values of the forget gate state, input gate state, output gate state, memory cell, and hidden state at time $t$ in the sequence, respectively. $\zeta$ and Tanh are the sigmoid function and hyperbolic tangent function, $W$ and $b$ are the weight matrix and bias vector, and $\otimes$ denotes the element-wise product. 

\nomenclature[P]{\(f\)}{Forget gate's activation vector}
\nomenclature[P]{\(i\)}{Input/update gate's activation vector}
\nomenclature[P]{\(o\)}{Output gate's activation vector}
\nomenclature[P]{\(c\)}{Cell state vector}
\nomenclature[P]{\(h\)}{Hidden state vector}
\nomenclature[P]{\(W\)}{Weight matrix}
\nomenclature[P]{\(b\)}{Bias vector}
\nomenclature[P]{\(\zeta\)}{Sigmoid function}
\nomenclature[P]{\(\text{Tanh}\)}{Hyperbolic tangent function}
\nomenclature[P]{\(\otimes\)}{Element-wise product}

\begin{figure*}[t!]
\begin{centering}
\includegraphics[]{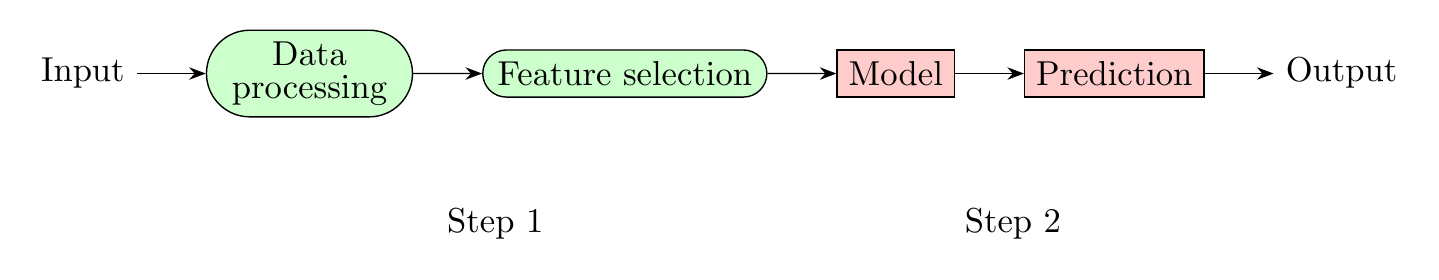}
 \caption{\centering The flowchart of a two-step hybrid model. The green nodes stand for the first step, and the red nodes stand for the second step.}\label{f4}
 \end{centering}
\end{figure*}

\begin{figure*}[t!]
\begin{centering}
\includegraphics[]{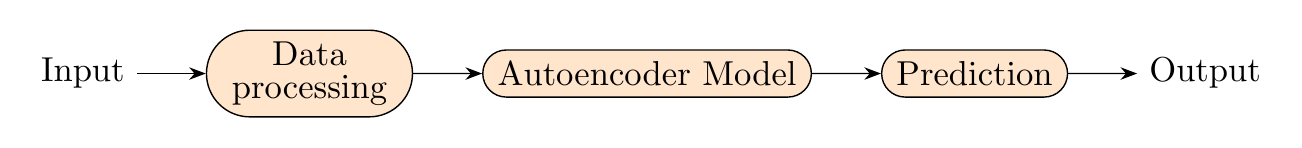}
\caption{\centering The flowchart of an autoencoder hybrid model. The orange nodes stand for the autoencoder process.}\label{f5}
\end{centering}
\end{figure*}

\begin{centering}
\begin{figure*}[t!]
\includegraphics[]{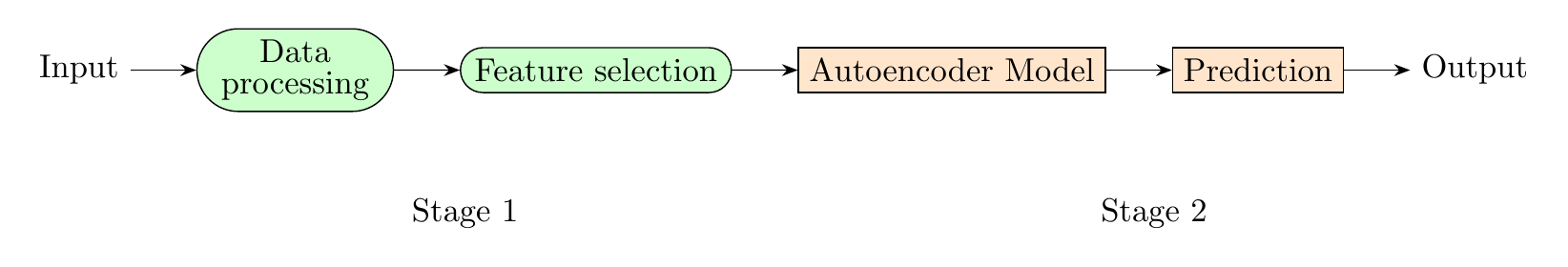}
\caption{\centering The flowchart of a two-stage hybrid model. The green nodes stand for the first stage, and the orange nodes stand for the second stage.}\label{f6}
\end{figure*}
\end{centering}

\subsection{Architectures of hybrid models} \label{S:4}
Typically, there are three hybrid architectures for EPF when working with LSTM. The first architecture consists of two steps, as shown in Figure \ref{f4}. The first step includes data processing and feature selection, and the second step contains training of the predictive models and making predictions.The second architecture can be referred to as an autoencoder model. Here, the input data will be turned into a compressed representation rather than specifically showing which features are selected, as shown in Figure \ref{f5}. The third combines the two aforementioned architectures, and it is referred to as two-stage feature selection. In this architecture, the explanatory variables will be selected by some feature selection method in the first stage. The selected features will then become the input for the autoencoder models in the second stage. Figure \ref{f6} shows this architecture.

\subsection{Feature selection methods} \label{S:5}
\subsubsection{PC}
The PC coefficient is a statistic used to measure the linear relationship between two data samples. Given two variables $(X, Y)$, the formula of the PC coefficient $\rho$ is given by the following:

\begin{equation}
\rho(X, Y) = \frac{cov(X,Y)}{\sigma_X \sigma_Y}
\end{equation}
\noindent where cov is the covariance, $\sigma_X$ is the standard deviation of $X$, and $\sigma_Y$  is the standard deviation of $Y$. 

\nomenclature[P]{\(\rho\)}{PC coefficient}
\nomenclature[P]{\(\sigma_X\)}{Standard deviation of $X$}
\nomenclature[P]{\(\sigma_Y\)}{Standard deviation of $Y$}
\nomenclature[P]{\(cov\)}{Covariance}

\subsubsection{PSO-ELM and GA-ELM}
PSO-ELM and GA-ELM are wrapper-based hybrid methods. ELM is a single hidden layer feedforward neural network. Its fast training \cite{HUANG2006489} contributes to the popularity of its employment as a predictive model in wrapper-based feature selection \cite{Saraswathi2011452, CHYZHYK201473, AHILA201523}. The output of ELM is calculated as follows:

\begin{equation}
F_L(x) =  \sum_{i=1}^Lw^2_i\phi(w^1_i x_j+b_i), j = 1, ..., N
\end{equation}
\nomenclature[P]{\(w^1\)}{Weight vector between the input and the hidden layer}
\nomenclature[P]{\(w^2\)}{Weight vector between the hidden layer and the output}
\nomenclature[P]{\(\phi\)}{Activation function}

\noindent where $L$ is the number of hidden units, $N$ is the number of training samples, $w^2$ is the weight vector between the hidden layer and the output, $w^1$ is the weight vector between the input and the hidden layer, $\phi(*)$ denotes an activation function, $b$ is a bias vector, and $x$ is the input vector.

PSO and GA are different types of optimisation algorithms, which provide the subsets of features as the input to the ELM to detect the optimal feature selection. The basic idea of PSO is that a swarm of particles moves through the search space. The movement of each particle is guided by its own known best-position and the entire swarm's known best position. PSO performs the search for the optimum by iteratively updating the velocities of the particles in the swarm \cite{Zhang2015}. The GA is a search metaheuristic that was inspired by Darwin’s theory of natural selection. In general, GAs will search for the optimal solution from a set of possible solutions, called a population. A solution is referred to as a chromosome or an individual. These chromosomes evolve over a number of generations by recombination (cross-over) and mutation \cite{Whitley1994}.  The detailed introduction of the methods can be found in Appendix A.1 and A.2.

\subsubsection{RFE-SVR}
RFE-SVR is another wrapper-based feature selection method. The core idea of this algorithm is to search for the best subset of features by starting with all features and discarding the less important features. In particular, the RFE algorithm operates with SVR to perform feature selection and regression simultaneously. SVR performs well in  high
dimensionality space \cite{NIPS1996_d3890178}. The detailed explanation of SVR is in Appendix A.3.

\subsubsection{Lasso regression}
The Lasso regression aims to increase the prediction accuracy of regression models by adding a penalty $\lambda \sum_{j=1}^n|\beta_j|$ to the loss function. This means that instead of minimising a loss function, $\sum_{i=1}^m(y_i-\sum_{j=1}^nx_{ij}\beta_j)^2$, the loss function becomes $\sum_{i=1}^m(y_i-\sum_{j=1}^nx_{ij}\beta_j)^2 + \lambda \sum_{j=1}^n|\beta_j|$, where $y$ is the vector of the dependent variable, $x$ denotes independent variables, the $\beta$ are the corresponding coefficients. The algorithm has the advantage that it shrinks some of the less critical coefficients of features to zero. Therefore, it removes less relevant features. 

\nomenclature[P]{\(\lambda\)}{Tuning parameter}
\nomenclature[P]{\(\beta\)}{Coefficients of independent variables}
\nomenclature[P]{\(x\)}{Input variable}
\nomenclature[P]{\(y\)}{Output variable}

\subsection{Autoencoder Model}\label{S:6}
\begin{figure}
  \centering
\includegraphics[width=0.45\textwidth]{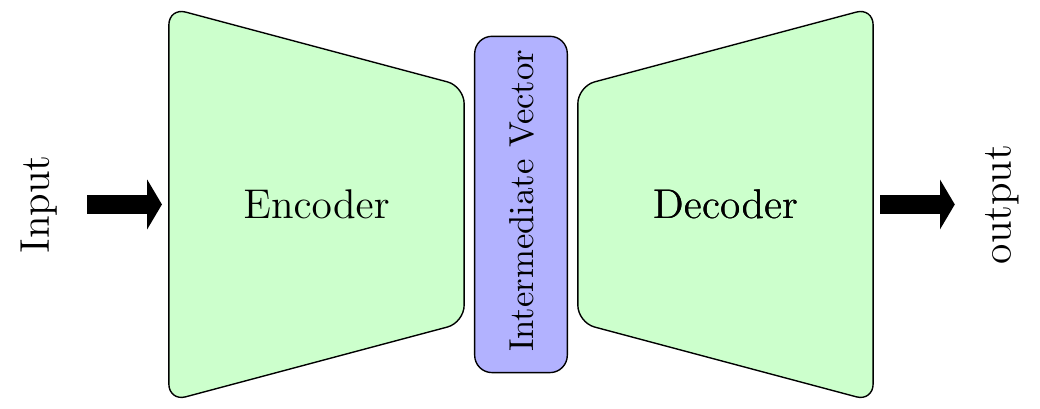}
\caption{\centering Structure of the Encoder-Decoder model.}\label{f11}
\end{figure}

An autoencoder is typically a neural network that aims to filter and compress the representation of its input, which consists of two components: an encoder and a decoder, shown in Figure \ref{f11}. The encoder typically accepts a set of input data and compresses the information into an intermediate vector. The decoder is typically a predictive model. In our case, a decoder is an LSTM network, and the encoders are LSTM, CNN, and convolutional layers, as described in the following.

\subsubsection{LSTM-LSTM Encoder-Decoder model}
In an LSTM-LSTM Encoder-Decoder model, an LSTM model is used as the encoder to process the raw input time series and to transform it into an intermediate vector. LSTM is capable of extracting the complex dynamic information within the temporal input series and filtering useful information from long input series via internal memory.

\subsubsection{CNN-LSTM Encoder-Decoder model}
In a CNN-LSTM Encoder-Decoder model, a CNN is the encoder to filter the input data. CNNs were originally and successfully used to process the image input data in image recognition tasks \cite{Szegedy2015} or the sequence of input data in natural language processing problems \cite{Sutskever2014}. The convolutional layers are usually followed by a pooling layer, which extracts information from the convolved features and produces a lower dimensional output. Then, the output values are flattened into a long intermediate vector representation. 

\subsubsection{Convolutional LSTM Encoder-Decoder model}
The computational mechanism of the convolutional LSTM  (ConvLSTM) is similar to that of CNN-LSTM \cite{SHI2015}. Unlike the CNN-LSTM, where the CNN model generates the input for the LSTM model, in the ConvLSTM model, the LSTM neural network processes the extracted information directly from preceding convolutional layers.

\section{Experimental details}
In this section, we introduce the concepts and methods employed in the process of training and evaluation of the constructed models. This section includes five parts. First, statistical performance measures are specified for evaluation and comparison. Second, theory and preliminary work for training models are explained. Third, configuration parameters are set for feature selection algorithms and LSTM-based models. Fourth and fifth, the benchmark model and feature explanation method used in the empirical study are introduced. They are necessary preparations for conducting the experiments.

\subsection{Statistical performance measures}

\subsubsection{Evaluation metrics}
In this paper, we employ several indicators to evaluate the accuracy of predictions: the mean absolute error (MAE), the root mean squared error (RMSE), the mean absolute percentage error (MAPE), and the symmetric mean absolute percentage error (SMAPE) as the model estimator. They are commonly adopted in EPF research \cite{Weron2006}. Given a predicted output vector, $\hat{y_k}=[\hat{y_1},..,\hat{y_N}]$, and a real output vector, $y_k=[y_1,..,y_N]$, the MAE, RMSE, MAPE, and SMAPE can be calculated as follows: 

\begin{equation}
\text{MAE} = \frac{1}{N}\sum_{k=1}^{N}{|y_k - \hat{y_k}|}
\end{equation}

\begin{equation}
\text{RMSE} = \sqrt{\frac{1}{N}\sum_{k=1}^{N}{(y_k - \hat{y_k})^2}}
\end{equation}

\begin{equation}
\text{MAPE} = \frac{100}{N}\sum_{k=1}^{N}\abs{\frac{y_k-\hat{y_k}}{y_k}}
\end{equation}

\begin{equation}
\text{SMAPE} = \frac{100}{N}\sum_{k=1}^{N}\frac{|y_k-\hat{y_k}|}{(|y_k|+|\hat{y_k}|)/2}
\end{equation}

\nomenclature[P]{\(\hat{y}\)}{Predicted output}

\subsubsection{Diebold-Mariano test}
The metrics for assessing the forecasting accuracy mentioned above cannot guarantee that the observed difference from two predictive models is statistically significant. In this context, the Diebold-Mariano (DM) test is typically used for evaluating the performance of two models \cite{Diebold1995134, HARVEY1997281}. Given the actual values of a time series $[{y_t; t = 1,..., T}]$, two forecasts from two models, $[{\hat{y}_{1t}; t = 1,..., T}]$ and $[{\hat{y}_{2t}; t = 1,..., T}]$, and the associated forecast errors, $e_{1t} =\hat{y}_{1t} - {y_t }$ and $e_{2t} =\hat{y}_{2t} - {y_t }$, the DM test defines the loss differential between the two forecasts by the following:
\begin{equation}
d_t^{F1,F2} = g(e_{1t}) - g(e_{2t})
\end{equation}
\noindent where $g(*)$ stands for loss function. In a one-sided DM test, the hypotheses is the following:
\begin{equation}
\begin{split}
H_0: \mathbb{E}{(d_t^{F1,F2})} \geq 0,\\
H_1: \mathbb{E}{(d_t^{F1,F2})} < 0.
\end{split}
\end{equation}
\noindent A one-sided DM test is used to detect whether F2 is better than F1. If $H_0$ is rejected, the test suggests that the accuracy of F1 is, statistically, significantly better than F2.
\noindent The complementary one-sided DM test can be expressed as follows:
\begin{equation}
\begin{split}
H_0: \mathbb{E}{(d_t^{F1,F2})} \leq 0,\\
H_1: \mathbb{E}{(d_t^{F1,F2})} > 0.
\end{split}
\end{equation}
If $H_0$ is rejected, the test suggests that the accuracy of F2 is, statistically, significantly better than F1. In this study, we employ a one-sided DM test to assess the forecasting performance of the proposed models. We chose $d_t^{F1,F2} = |e_{1t}| - |e_{2t}|$ as the loss differential.

\nomenclature[P]{\(d^{F1,F2} \)}{Loss differential}
\nomenclature[P]{\(g(*) \)}{Loss function}
\nomenclature[P]{\(e \)}{Forecast error}

\subsection{Model training}
\subsubsection{Walk forward nested cross-validation}
\begin{figure}
  \centering
\includegraphics[width=0.5\textwidth]{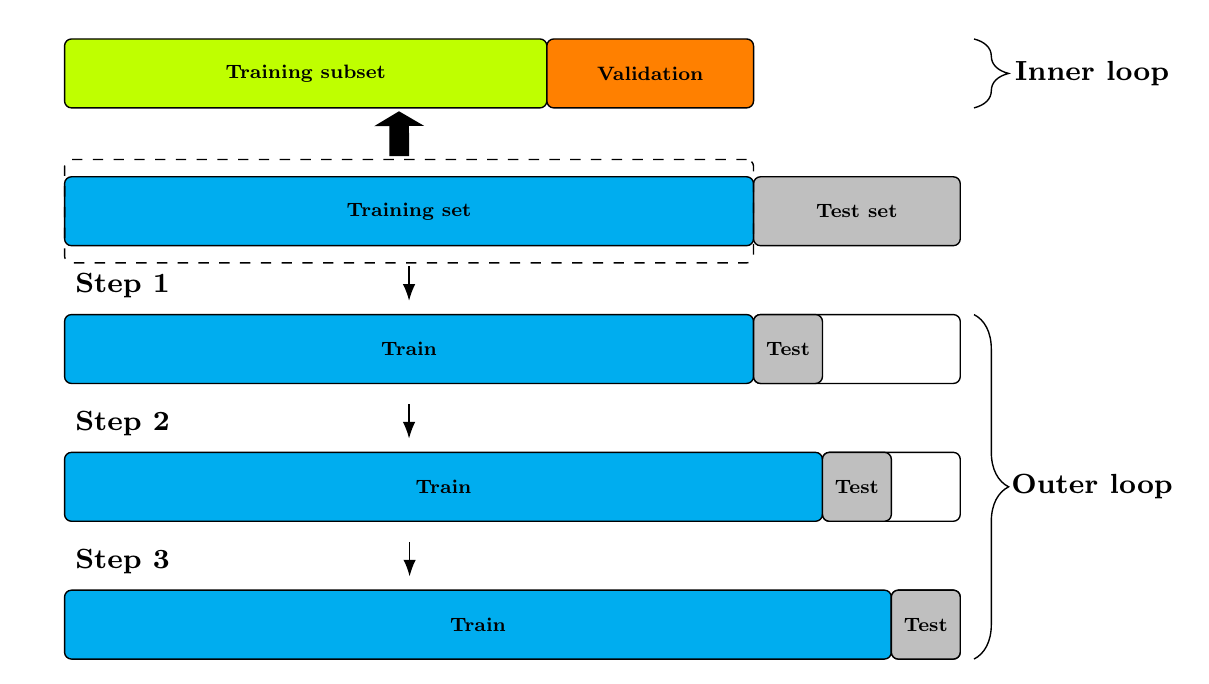}
\caption{\centering Walk forward nested cross-validation.}\label{f12}
\end{figure}

To avoid over-fitting, it is common to include a validation set to evaluate the generalisation ability of the training model. The cross-validation is referred to as a method for tuning the hyperparameters and producing robust measurements of model performance. In \cite{Varma2006}, a nested cross-validation procedure was introduced, which considerably reduced the bias and provided an almost unbiased estimate of the true error. Because new observations become available over time, in time series modelling, we implemented a walk forward nested cross-validation in which the forecast rolls forward in time. More specifically, we successively considered each day as the test set and assigned all previous data to the training set (Outer loop). The training set is split into a training subset and a validation set. The validation set data comes chronologically after the training subset (Inner loop). Walk forward validation involves moving along the time series one time step at a time. The process requires multiple models to be trained and evaluated, but the additional computational cost will provide a more robust estimate of the expected performance of the predictive model on unseen data. It is shown in Figure \ref{f12}. 

\subsubsection{Data division}
We divided the whole database into two subsets: a training set and a test set. The training set includes a training subset and a validation subset, as shown in the dashed box in Figure \ref{f12}. We initially apportioned the data set into training, validation, and test sets, with an 80-10-10 split. The magnitude of the test and validation set is anchored during the walk-forward test. 

\subsubsection{Data processing}
For neural network model training, the input data is usually normalised to the intervals [0,1]. This is not only done because the normalised data will require less time to train, but the prediction performance will also increase. In addition, we linearly interpolate the missing data and eliminate duplicates due to daylight saving.

\subsubsection{Ten experiments}
Training algorithms for deep learning models have usually required the initialisation of the weights of neural networks from which to begin the iterative training \cite{Goodfellow2016}. The random initial conditions for an LSTM network can result in different performances each time a given configuration is trained. Thus, we employed ten experiments for each model to reduce the impact of the variability on performance evaluation. Models were evaluated after taking the average of the experiments.

\subsection{Model configuration parameters}
\subsubsection{Parameters of feature selection}
The feature selection stopping criterion varies by algorithm, which is controlled by the parameters of models. The applied configuration of PSO was [$c_1$: 0.5, $c_2$: 0.3, $\omega$: 0.7], and the stop condition is satisfied after 10,000 iterations. For GA, the crossover possibility and mutation possibility were set to 0.5 and 0.2, respectively. The population size was 100, and the maximum number of generations was 10,000. On the basis of the predictive ELM, the amount of the selected features by PSO-ELM and GA-ELM was automatically set to 30. For the sake of input consistency, the magnitude of the selected features of the rest of the models was set to 30 as well. For the PC method, we ranked all features attributable to the correlation coefficients and selected the first 30 features. In terms of RFE-SVR, we ranked features by importance, discarded the least important features, and refit the model until 30 features remained. The regularisation parameter, $\lambda$, in Lasso regression was 0.02.

\subsubsection{Network hyperparameters}
 Our study aimed to investigate the applications and impacts of different types of feature selection methods in a predictive LSTM architecture. We used a coherent configuration of a specific LSTM model for comparison and did not perform an extensive hyperparameter optimisation to search for the optimal configuration. After an inexhaustive grid search, we constructed our prediction model from an LSTM model with a single hidden layer of 300 units, followed by a fully connected dense layer with 100 neurons that preceded the output layer. The LSTM encoder has a hidden layer with 300 units. In the CNN-LSTM Encoder-Decoder model, the CNN encoder has two convolutional layers, with 96 units to amplify any salient features, followed by a max-pooling layer. In the ConvLSTM Encoder-Decoder model, the encoder is a convolutional layer with 64 units. The input sequence length is 14 days (2 weeks, commonly used in EPF). The optimiser is the Adam algorithm, and the loss function is Mean Squared Error (MSE).

\subsection{Benchmark model}
Among the traditional methods, the statistical models perform best for EPF. Thus, we select the Nonlinear AutoRegressive Moving Average with eXogenous Input (NARMAX) model as the benchmark (trained with the optimal structure) for our case study. This statistical model is widely used in energy price forecasting to handle multiple nonlinear inputs \cite{Johannesen2019, McHugh2019}. The equation is represented as the following:
\begin{equation}
\begin{split}
y ( t ) &= F ^ { \ell } \left[ y ( t - 1 ) , \ldots , y \left( t - N _ { y } \right), x ( t ) , \ldots , x \left( t - N _ { x } \right), \right. \\ &\varepsilon ( t - 1 ) , \ldots \varepsilon \left( t - N _ { \varepsilon } \right) ] + \varepsilon ( t )
\end{split}
\end{equation}

\nomenclature[P]{\(\varepsilon  \)}{Error variable}

\noindent where $x(t)$ is the input and $y(t)$ is the output time-series; $\varepsilon ( t )$ is the uncertainties and possible noise; $N_u$, $N_y$, and $N_\varepsilon$ are the input, output, and prediction error lags, respectively; and $F ^ { \ell }$ is a nonlinear function.
\subsection{Feature explanation method}
In this study, we used SHapley Additive exPlanations (SHAP) values to interpret the impact of certain values of a given feature on the expected price prediction. SHAP\footnote{The Python package SHAP is available at \url{https://github.com/slundberg/shap}} is a theoretic game method to explain the output of machine learning models \cite{Lundberg2020, janzing2019feature, sundararajan2020shapley}. The Shapley value is used to assess the feature relevance relative to the expectation of the output \cite{lundberg2017unified}. In particular, a Kernel SHAP is used for explaining an optimal SVR model obtained by grid-search on the dataset.

\section{Results}\label{S:8}
In this section, we report the empirical results obtained by the application of the introduced models. For similarity of presentation, the list of models and their acronyms are shown in Table \ref{t3}.
\begin{centering}
\begin{table}[!t]
\caption{\\The proposed models.}\label{t3}
\centering
 \resizebox{0.48\textwidth}{!}{
 \begin{tabular}{l l l} 
\toprule
Mode & Category&Model Explanation\\
\midrule
 M0 &Benchmark& NARMAX model \\
 M1 &Filter method& PC-LSTM model  \\
 M2 &Wrapper method& PSO-ELM-LSTM model  \\
 M3 &Wrapper method& GA-ELM-LSTM model  \\
 M4 &Wrapper method& RFE-SVR-LSTM model  \\
 M5 &Embedded method& LASSO-LSTM model  \\
 M6 &Autoencoder method& LSTM-LSTM Encoder-Decoder model \\
 M7 &Autoencoder method& CNN-LSTM Encoder-Decoder model  \\
 M8 &Autoencoder method& CovLSTM Encoder-Decoder model  \\
 M9 &Two-stage method& PC-LSTM-LSTM Encoder-Decoder model  \\
 M10 &Two-stage method& PSO-ELM-LSTM-LSTM Encoder-Decoder model  \\
 M11 &Two-stage method& GA-ELM-LSTM-LSTM Encoder-Decoder model  \\
 M12 &Two-stage method& RFE-SVR-LSTM-LSTM Encoder-Decoder model  \\
 M13 &Two-stage method&  LASSO-LSTM-LSTM Encoder-Decoder model  \\
\bottomrule
\end{tabular}} 
\end{table}
\end{centering}
\subsection{Analysis of empirical results}
The results of the feature selection are shown in Table \ref{t_s}. Overall, it can be observed that different selection mechanisms lead to different selections. From Table \ref{t_s}, we can see that M1 selects all the day-ahead prices. It is not surprising that the day-ahead prices from different bidding areas are more relevant to the Nord Pool system price than other feature variables. However, the over-selection results in information redundancies. Some researchers have recognised that the redundancy among features decreases the model's performance \cite{Peng20051226, Yu20041532, Langley1994f}. Compared to M1, the wrapper-based methods, M2 and M3, eliminate several price variables rather than the other categories of variables. It is worth noticing the two methods does not select the lag system price (F1), which is commonly used in time series, given their short-term autoregressive nature. As introduced in section 3.3.2, PSO-ELM  and  GA-ELM are widely used in research. However, the optimisation methods, such as PSO and GA, have the problem of trapping in local optima. Although re-setting and experimenting can increase the chance of avoiding traps, when dealing with high dimensional data sets, the optimisation methods still cannot guarantee they will find a global optimum solution, and they are not suitable for all cases \cite{Jamian2014}. The straightforward concept and fast computation of ELM contributed to its widespread application in an exhaustive grid search, but it does not consider the sequential relationships in time series data, as with other traditional neural networks. These could be the reasons why the two methods eliminate the lag system price as an input. The other wrapper model M4, selects various types of features, and eliminates less price features compared to M2 and M3. It is interesting to note that M4 does not pick up any features from the cross-border flow deviation. For the Lasso regression method, M5, we can summarise that it diversely chooses features such as M4 but puts more emphasis on electricity transmission.  

\xentrystretch{-0.15}
\begin{center}
\topcaption{\centering The results of feature selection.\label{t_s}}
\tablefirsthead{\toprule \multirow{2}{*}{Feature}&\multicolumn{5}{c}{Feature selection model}\\ \cline{2-6}&M1&M2&M3&M4&M5\\ }
\tablehead{%
\multicolumn{6}{c}%
{{\bfseries  Continued from previous column}} \\
\toprule
\multirow{2}{*}{Feature}&\multicolumn{5}{c}{Feature selection model}\\ \cline{2-6}&M1&M2&M3&M4&M5\\ \midrule}
\tabletail{%
\midrule \multicolumn{6}{r}{{Continued on next column}} \\ \midrule}
\tablelasttail{%
\multicolumn{6}{c}{{Note: \cmark \ denotes that the feature is selected;}} \\
\multicolumn{6}{c}{{\xmark  \ denotes that the feature is not selected.}} \\}
\begin{xtabular*}{0.36\textwidth}{p{0.048\textwidth} p{0.035\textwidth}p{0.035\textwidth} p{0.035\textwidth} p{0.035\textwidth} p{0.035\textwidth}} 
\toprule
F1&\cmark&\xmark&\xmark&\cmark&\cmark\\\shrinkheight{-5\normalbaselineskip}
F2&\cmark&\cmark&\cmark&\cmark&\cmark\\
F3&\cmark&\xmark&\xmark&\cmark&\cmark\\
F4&\cmark&\cmark&\xmark&\cmark&\xmark\\
F5&\cmark&\xmark&\xmark&\xmark&\xmark\\
F6&\cmark&\xmark&\xmark&\xmark&\xmark\\
F7&\cmark&\xmark&\xmark&\xmark&\xmark\\
F8&\cmark&\cmark&\xmark&\xmark&\xmark\\
F9&\cmark&\xmark&\xmark&\cmark&\xmark\\
F10&\cmark&\xmark&\xmark&\cmark&\cmark\\
F11&\cmark&\cmark&\xmark&\cmark&\cmark\\
F12&\cmark&\xmark&\cmark&\cmark&\cmark\\
F13&\cmark&\cmark&\cmark&\cmark&\cmark\\
F14&\cmark&\xmark&\cmark&\xmark&\xmark\\
F15&\cmark&\xmark&\xmark&\xmark&\xmark\\
F16&\cmark&\xmark&\xmark&\xmark&\xmark\\
F17&\cmark&\xmark&\cmark&\xmark&\xmark\\
F18&\cmark&\cmark&\xmark&\cmark&\cmark\\
F19&\xmark&\cmark&\xmark&\cmark&\xmark\\
F20&\xmark&\cmark&\cmark&\cmark&\xmark\\
F21&\xmark&\xmark&\cmark&\xmark&\xmark\\
F22&\xmark&\cmark&\cmark&\cmark&\xmark\\
F23&\xmark&\cmark&\xmark&\cmark&\xmark\\
F24&\cmark&\cmark&\cmark&\xmark&\xmark\\
F25&\xmark&\xmark&\xmark&\cmark&\cmark\\
F26&\xmark&\xmark&\cmark&\cmark&\cmark\\
F27&\cmark&\cmark&\cmark&\xmark&\cmark\\
F28&\xmark&\cmark&\cmark&\xmark&\xmark\\
F29&\xmark&\xmark&\xmark&\cmark&\cmark\\
F30&\xmark&\cmark&\xmark&\xmark&\cmark\\
F31&\xmark&\cmark&\xmark&\cmark&\xmark\\
F32&\xmark&\cmark&\cmark&\cmark&\cmark\\
F33&\cmark&\cmark&\xmark&\xmark&\xmark\\
F34&\cmark&\xmark&\cmark&\cmark&\cmark\\
F35&\xmark&\xmark&\xmark&\cmark&\cmark\\
F36&\xmark&\cmark&\cmark&\cmark&\cmark\\
F37&\xmark&\xmark&\xmark&\cmark&\xmark\\
F38&\cmark&\cmark&\cmark&\cmark&\xmark\\
F39&\cmark&\cmark&\cmark&\cmark&\cmark\\
F40&\xmark&\xmark&\cmark&\xmark&\xmark\\
F41&\cmark&\cmark&\xmark&\cmark&\cmark\\
F42&\xmark&\xmark&\xmark&\xmark&\xmark\\
F43&\cmark&\xmark&\cmark&\cmark&\xmark\\
F44&\cmark&\cmark&\cmark&\cmark&\cmark\\
F45&\cmark&\xmark&\cmark&\xmark&\xmark\\
F46&\xmark&\cmark&\xmark&\xmark&\cmark\\
F47&\xmark&\xmark&\xmark&\xmark&\xmark\\
F48&\xmark&\xmark&\xmark&\cmark&\cmark\\
F49&\xmark&\xmark&\cmark&\xmark&\cmark\\
F50&\xmark&\cmark&\cmark&\xmark&\cmark\\
F51&\xmark&\cmark&\xmark&\xmark&\cmark\\
F52&\xmark&\cmark&\cmark&\xmark&\xmark\\
F53&\xmark&\xmark&\xmark&\cmark&\cmark\\
F54&\xmark&\xmark&\cmark&\xmark&\cmark\\
F55&\xmark&\xmark&\xmark&\xmark&\xmark\\
F56&\xmark&\xmark&\cmark&\xmark&\xmark\\
F57&\xmark&\cmark&\cmark&\xmark&\xmark\\
F58&\xmark&\xmark&\cmark&\xmark&\cmark\\
F59&\cmark&\cmark&\cmark&\xmark&\cmark\\
F60&\cmark&\cmark&\cmark&\xmark&\xmark\\
F61&\xmark&\cmark&\xmark&\xmark&\cmark\\
F62&\xmark&\xmark&\xmark&\xmark&\xmark\\
\bottomrule
\end{xtabular*}
\end{center}
\begin{figure}[ht!]
\centering
\includegraphics[width=0.45\textwidth, height = 4.5cm]{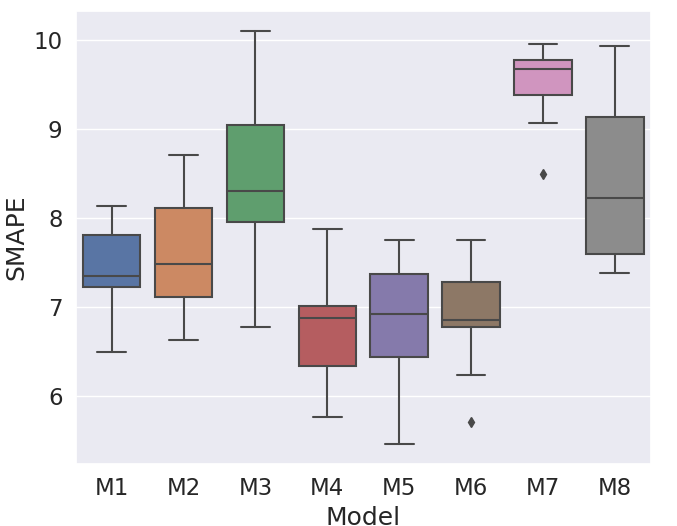}
\caption{\centering The SMAPEs of 10 experiments for M1, M2, M3, M4, M5, M6, M7, and M8.}\label{f_smape_10}
\end{figure}

\begin{table*}[ht!]
\caption{\\The results of the one-sided DM test.}\label{t_dm_all}
 \begin{threeparttable}
 \begin{tabular*}{\textwidth}{l P{1.4cm} P{1.4cm} P{1.4cm} P{1.4cm}P{1.4cm} P{1.4cm} P{1.4cm}P{1.4cm}P{1.4cm}} 
\toprule
\diagbox{\small{F1}}{\small{F2}}&M0&M1&M2 &M3&M4&M5&M6&M7&M8\\
\midrule
M0&& 5.83*** &4.99*** &3.90*** &6.65*** &7.09*** &6.60*** &4.61*** &4.27***\\
M1&-5.83***  & &-1.15 &-1.67* &2.31** &2.98*** &1.59\# &-0.16 &0.88\\
M2&-4.99*** &1.15  & &-0.91 &2.70*** &3.55*** &3.84*** &0.84 &-0.20\\
M3&-3.90*** &1.67* &0.91  & &3.19*** &4.01*** &4.31*** &1.33 &-0.97\\
M4&-6.65*** &-2.31** &-2.70*** &-3.19***  & &1.12 &-0.23 &-1.64\# &2.57**\\
M5&-7.09*** &-2.98*** &-3.55*** &-4.01*** &-1.12  & &-0.97 &-2.33** &3.19***\\
M6&-6.60*** &-1.59\# &-3.84*** &-4.31*** &0.23 &0.97  & &-1.44 &3.54***\\
M7&-4.61*** &0.16 &-0.84 &-1.33 &1.64\# &2.33** &1.44  & &0.66\\
M8&-4.27*** &-0.88 &0.20 &0.97 &-2.57** &-3.19*** &-3.54*** &-0.66  & \\
\bottomrule
\end{tabular*}
 \begin{tablenotes}
      \small
      \item Note: ***, ** ,* and \# denote 1\%, 5\%, 10\%, and 15\% significance levels, respectively. The positive sign of the DM value indicates that F2 is better F1. The negative sign of the DM value indicate that F1 is better F2.
    \end{tablenotes}
  \end{threeparttable}
\end{table*}

\begin{table}[h!]
\caption{\\The SMAPE of M0, M1, M2, M3, M4, M5, M6, M7, and M8.}\label{t_smape_all}
 \resizebox{0.5\textwidth}{!}{
 \begin{tabular}{l c c c c c c c c c} 
\toprule
Model&M0&M1&M2&M3&M4&M5&M6&M7&M8\\
\midrule
SMAPE&10.07&6.25&6.58&7.06&5.29&4.89&5.20&6.14&6.53\\
\bottomrule
\end{tabular}}
\end{table}
\begin{table}[t!]
\caption{\\The results of the one-sided DM test when comparing two-step models (F1) and two-stage models (F2).}\label{t_two_stage}
 \resizebox{0.5\textwidth}{!}{
 \begin{threeparttable}
 \begin{tabular}{c c c c c c} 
\toprule
F1 &M1 &M2&M3&M4&M5\\
\midrule
&-0.6039&0.2222&2.4556 ***&2.4524 ***&-1.7053 *** \\
\midrule
 F2&M9 &M10&M11&M12&M13\\
\bottomrule
\end{tabular}
 \begin{tablenotes}
      \small
      \item Note: ***, ** ,* and \# denote 1\%, 5\%, 10\%, and 15\% significance levels, respectively. The positive sign: F2 is better F1. The negative sign: F1 is better F2.
    \end{tablenotes}
  \end{threeparttable}}
\end{table}

To evaluate the statistical significance in the difference of predictive accuracy, one-side DM tests, as defined in section 4.1.2, were applied, and the results are shown in Table \ref{t_dm_all}. Table \ref{t_smape_all} exhibits the performance comparison of all the models in terms of SMAPE. As expected, the proposed LSTM models are overwhelmingly better than the benchmark statistical model, M0. The superior performance of deep learning models to statistical models has been recognised by numerous studies \cite{LAGO2018386, SOMU2021110591}. Moreover, the performances of LSTM models in ten experiments are depicted in Figure \ref{f_smape_10}, measured in terms of SMAPE. As seen in Figure \ref{f_smape_10}, of all the models, M4, M5, and M6 perform better than the others. The statistical details of the model performance in ten experiments are listed in Appendix Tables \ref{t11}, \ref{t12}, \ref{t13}, and \ref{t_smape_table}  by means of MAD, RMSE, MAPE, and SMAPE. Based on the analysis of feature selection, we conclude that M4 and M5, i.e., the minimum redundancy maximum relevance algorithms, perform better than the others. The results are consistent with the observations from other literature \cite{SHAO2017330, Radovic2017} that an elimination of the redundant and less relevant features increases the performance of models. In addition, some researchers have attempted to introduce the CNN-LSTM model and show its excellent performance in the energy field \cite{Kim2018, KIM201972}. However, we found that M6 performs better than M7 and M8. This means LSTM-LSTM is a better autoencoder structure than CNN-LSTM and ConvLSTM for EPF. The results are reasonable because convolutional neural networks (CNNs or ConvNets) were originally designed for image recognition and classification, while recurrent neural networks (LSTM) are for sequence and time series prediction.

\begin{figure}[t!]
\centering
\includegraphics[width=0.43\textwidth, height = 4.cm]{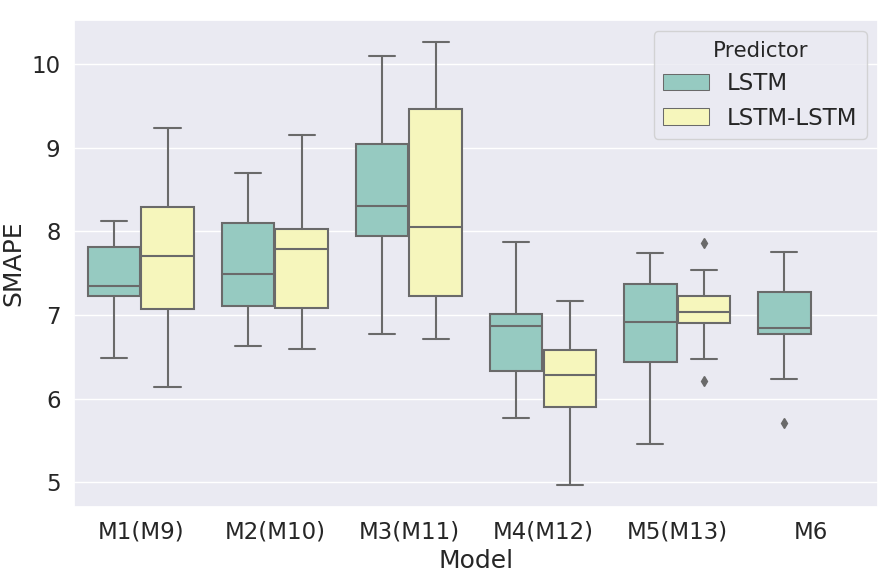}
\caption{\centering The comparison of SMAPEs between two-step LSTM models and two-stage LSTM-LSTM models.}\label{f_two_stage}
\end{figure}
We show the comparison of the SMAPE of the two-step and the two-stage models in Figure \ref{f_two_stage}. From this figure, it can be seen that M4 has been improved by applying LSTM-LSTM as predictors. We used a one-sided DM test to detect whether the two-stage LSTM models were statistically better than two-step LSTM-LSTM models. The results are shown in Table \ref{t_two_stage}. The superior features selection from M4 provides the possibility for autoencoder models to further process the selected features to obtain more meaningful information.

\begin{figure}[ht!]
\centering
\includegraphics[width=0.45\textwidth,height =4cm]{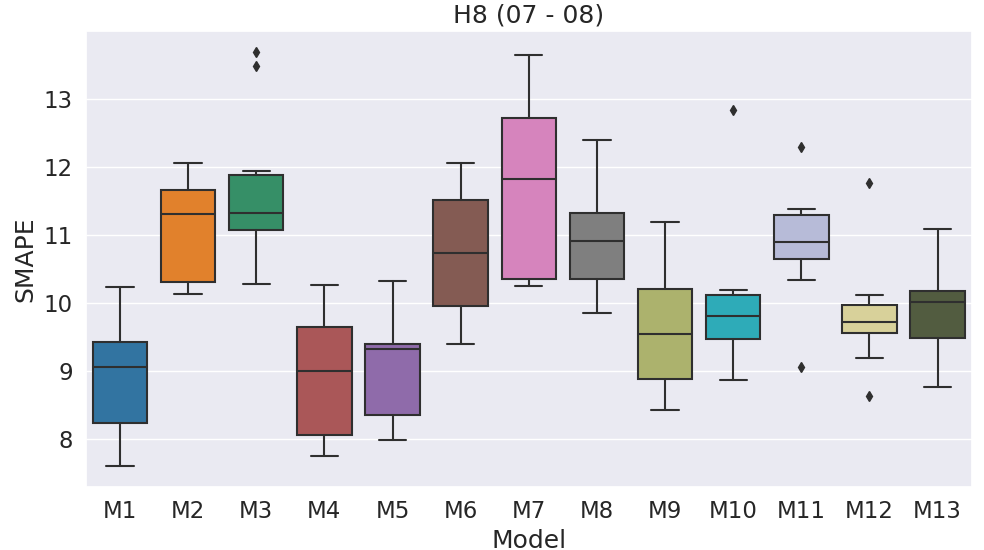}
\caption{\centering The SMAPEs of 10 experiments for predicting H8.}\label{f_h8}
\end{figure}

\begin{figure}[h!]
\centering
\includegraphics[width=0.45\textwidth,height =4cm]{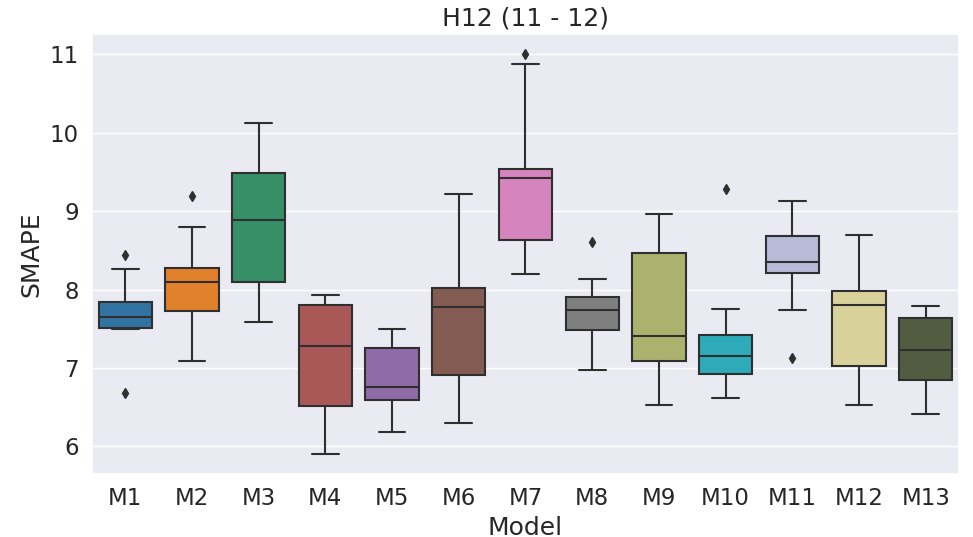}
\caption{\centering The SMAPEs of 10 experiments for predicting H12.}\label{f_h12}
\end{figure}

\begin{figure}[h!]
\centering
\includegraphics[width=0.45\textwidth,height =4cm]{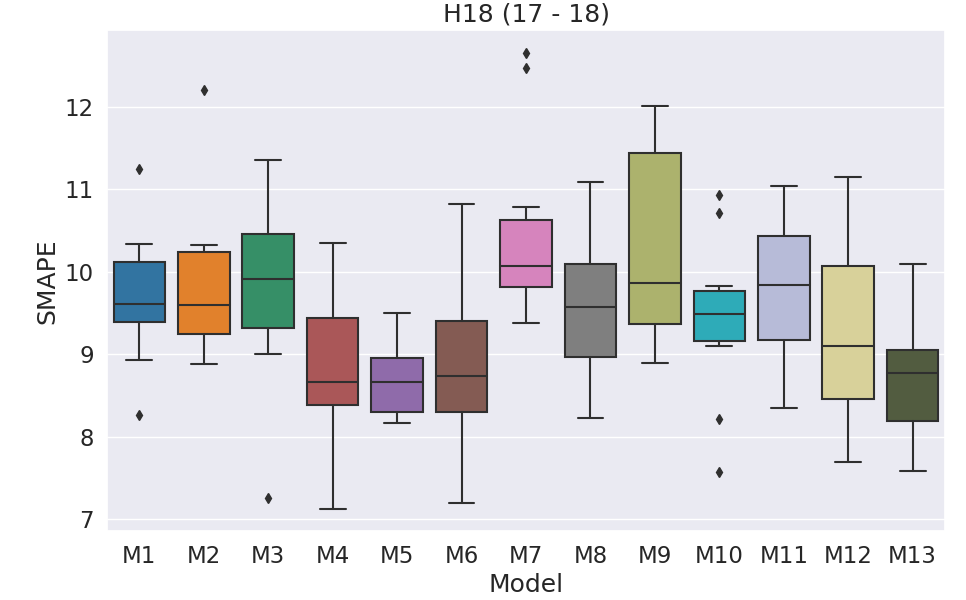}
\caption{\centering The SMAPEs of 10 experiments for predicting H18.}\label{f_h18}
\end{figure}

Additionally, we detected the forecasting performance for 24 hourly system prices. Figures \ref{f_h8}, \ref{f_h12}, and \ref{f_h18} show the results for the three peak hours: H8 (07 - 08), H12 (11 - 12) and H18 (17 - 18), respectively, measured in terms of SMAPE. We observed that the feature selections influence the forecasting accuracy and the models M4 and M5 are still relatively stable, performing better than other models. Indeed, the RFE-SVR and Lasso regression feature selection methods are applied to various areas in energy finance and achieve good performance for improving forecasting accuracy \cite{ Sultana2019259, Nawaz2020521, Brusaferri20191051, Leerbeck2020}.


\begin{centering}
\begin{figure*}[h!]
\centering
     \centering
     \begin{subfigure}[b]{0.4\textwidth}
         \centering
         \includegraphics[width=\textwidth]{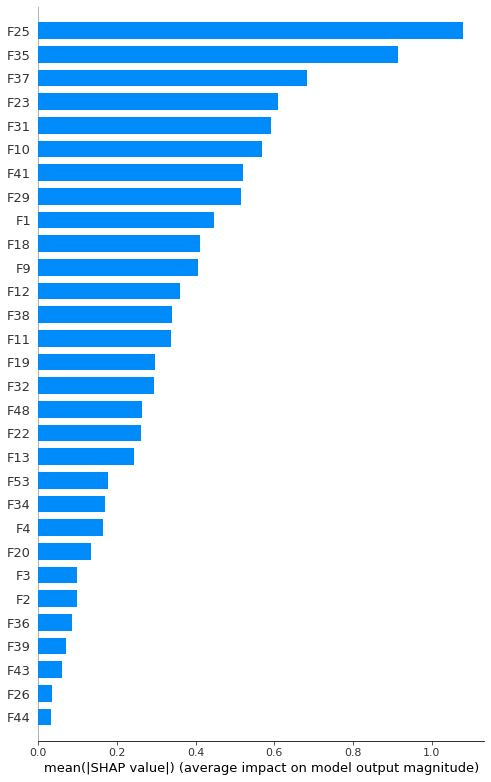}
         \caption{\centering{Features ranking}}
         \label{fig:r1}
     \end{subfigure}
     \begin{subfigure}[b]{0.4\textwidth}
         \centering
         \includegraphics[width=\textwidth]{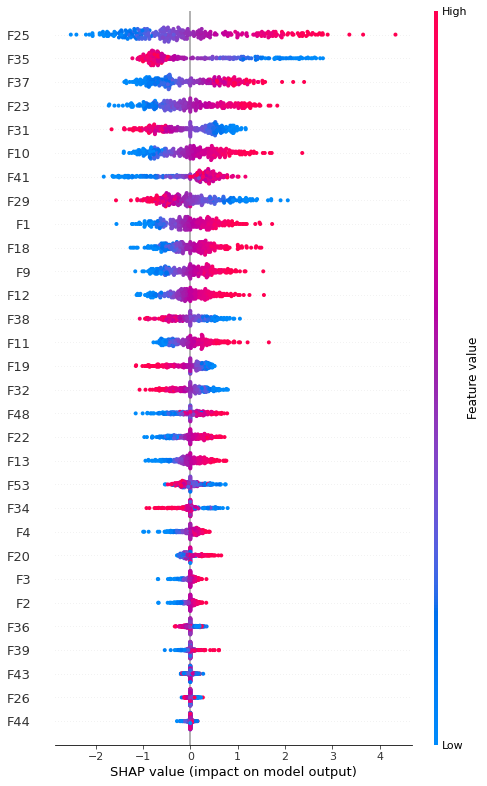}
         \caption{\centering{Features impact}}
         \label{fig:r2}
     \end{subfigure}
        \caption{The feature ranking and feature impact of the selected features of RFE-SVR. (a) Bar chart of the average SHAP value magnitude showing the importance of the features. (b) A set of beeswarm plots, where each dot corresponds to an individual day-ahead price. The dot’s position on the x-axis shows the impact that feature has on the model’s prediction for that price. Multiple dots landing at the same x position pile up to show density.}
        \label{fig:f_r}
\end{figure*}
\end{centering}
\subsection{Analysis of feature impact}
Figure \ref{fig:f_r} shows the ranking of features and their impacts on the predicted price in terms of the selected features of the RFE-SVR model (M4), which is the model with the best performance. From Figure \ref{fig:r1}, we can observe that the features from supply/demand sides are more important than the other features. In particular, production and consumption and their prognosis in the Nordic and German markets are prioritised by the model. The significant impact from the German market can be explained by the fact that the German market has the most electricity cables and the highest electricity exports to the Nordic market, as shown in Figures \ref{f1} and \ref{f_flow}. This indicates that it is critical to consider features from cross-border markets with increasing interconnections across Europe for EPF. Besides, electricity prices have more impact on EPF than the features from cross-border electricity trade. From Figure \ref{fig:r2}, we can see that the features have asymmetric predictive influence on electricity price.  For instance, the impact of DE consumption (F35) on EPF has a long-tail reaching to the right but not to the left. This indicates that German over-consumption can result in high Nordic electricity prices, but scarce consumption cannot significantly lower the price. 

\begin{centering}
\begin{figure*}[th!]
\centering
     \centering
     \begin{subfigure}[b]{0.32\textwidth}
         \centering
         \includegraphics[width=\textwidth,height = 4cm]{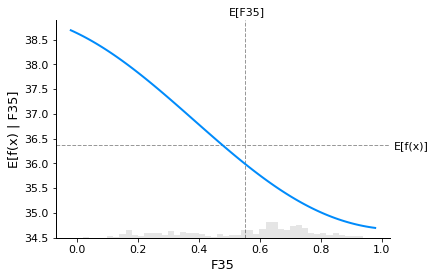}
         \caption{\centering{SHAP partial dependence plot of DE consumption}}
         \label{fig:dec0}
     \end{subfigure}
     \begin{subfigure}[b]{0.32\textwidth}
         \centering
         \includegraphics[width=\textwidth,height = 4cm]{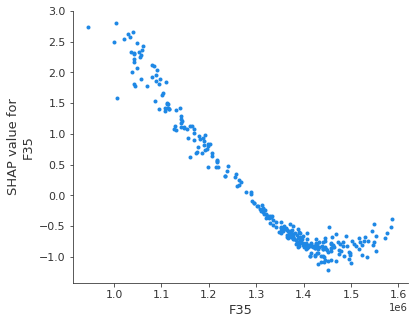}
         \caption{\centering{SHAP dependence plot of DE consumption}}
         \label{fig:dec1}
     \end{subfigure}
    \begin{subfigure}[b]{0.32\textwidth}
         \centering
         \includegraphics[width=\textwidth,height = 4cm]{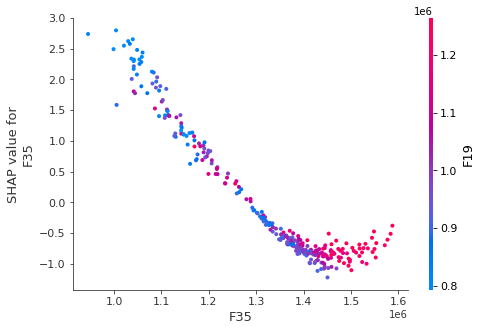}
         \caption{\centering{SHAP dependence plot of the interaction effect of DE consumption with the Nordic production (F19)}}
         \label{fig:dec2}
     \end{subfigure}
        \caption{The SHAP partial dependence (a) and dependence (b, c) plot of DE consumption (F35). $\mathbb{E}$[F35] is the expectation of the DE consumption, and  $\mathbb{E}$[f(x)] is the expectation of the Nordic price. The grey histogram in (a) shows the distribution of the feature in the test dataset. For (a), the x-axis is the normalised DE consumption. The x-axes in (b) and (c) are the real values of the consumption.}
        \label{fig:dec}
\end{figure*}
\end{centering}

Moreover, we detected the relations between different types of features and the predicted price with their dependence plots. Figure \ref{fig:dec0} demonstrates the negative association between DE consumption and its conditional expectation of the predicted price. If the DE consumption is high, then its value tends to revert to its expectation. Thus, the downward expectation of DE consumption will lead to the expected decline of the import demand from the Nordic market, which further decreases the expectation of the Nordic price. Figure \ref{fig:dec1} represents the change in predicted price as DE consumption changes. Vertical dispersion at a single value of DE consumption represents the interaction effects with other features. For example, the interaction effect of DE consumption with the Nordic production (F19) is shown in Figure \ref{fig:dec2}. The dependence plot highlights that the impact of DE consumption differs with different levels of the Nordic production. The results reveal that the Nordic price is less sensitive to the German power consumption when the Nordic electricity is oversupplied. It indicates, in such a case, that the information from the Nordic market rather than cross-border countries drives the price prediction.

\begin{figure}[h!]
\centering
\includegraphics[width =0.4\textwidth]{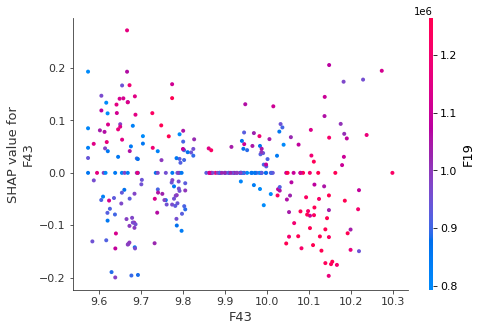}
\caption{The SHAP dependence plot of EUR/NOK (F43). The interaction effect of EUR/NOK with the Nordic production (F19).}\label{f_exchange}
\end{figure}

From Figure \ref{f_exchange}, we can see that the majority of the EUR/NOK exchange rates (F43) have no contribution to the prediction of the Nordic price (the y-axis value of the dots is zero). In addition, there is no obvious interaction effect of the Nordic productions and the exchange rates on the price. Thus, the predictive importance of exchange rate is extremely limited. 

\begin{figure}[h!]
\centering
\includegraphics[width =0.4\textwidth]{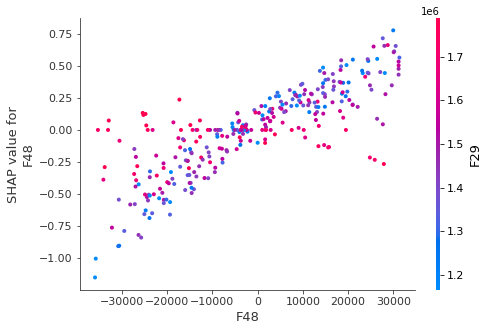}
\caption{The SHAP dependence plot of DK1 $\leftrightarrow$ DE flow (F48). The interaction effect of DK1 $\leftrightarrow$ DE flow with DE production prognosis (F29).}\label{fig:f_flow}
\end{figure}

From Figure \ref{fig:f_flow}, we can find that the predicted price is expected to increase when observing a high DK1$\rightarrow$DE cross-border electricity flow, indicating a relatively low current Nordic price. By contrast, a high flow from DE to DK1 implies that the Nordic price is relatively high and expected to decline. In addition, it can be seen via an interaction effect of the DK1 $\leftrightarrow$ DE flow with the DE production prognosis that the flow has less impact on the predicted price, with high expected production in Germany. The high production prognosis from cross-border countries will lead to a sharp decline in the expected cross-border transmission. Thus, the impact of the cross-border flow on the Nordic price formation on the following day will decrease significantly.

\begin{figure}[t!]
\centering
\includegraphics[width =0.4\textwidth]{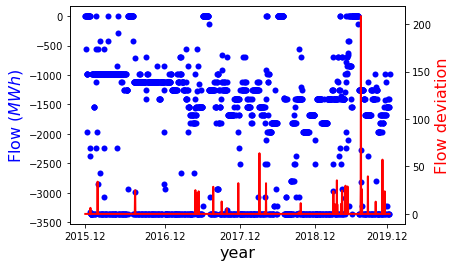}
\caption{FI $\leftrightarrow$ Russia flow (F54) (blue dots) versus FI $\leftrightarrow$ Russia flow deviation (F62) (red line).}\label{f_f_r}
\end{figure}

Last but not least, not all of the cross-border electricity flows and flow deviations are helpful for forecasting. The reason for this is that, in many cases, the flow capacity is fully occupied. The lack of variability results in their inability to provide useful information for forecasting. An example of the flow and flow deviations between FI and Russia can be seen in Figure \ref{f_f_r}. From Figure \ref{f_f_r}, it is evident that the majority of flow deviations are zero. The findings indicate that the non-selection of flow deviations from M4 is essential and reasonable. However, the capacity utilisation indicates the potential for more electrical power transmission across the Europe-wide market which would increase the overall socio-economic benefits.

\subsection{Discussion of practical importance}
The obtained results in the empirical study show that the LSTM-based hybrid models with various features from cross-border markets have considerably accurate prediction results for electricity price. An accurate prediction can be highly beneficial for the electricity market participants in practice. A power market firm that is capable of forecasting the volatile electricity price with a reasonable level of accuracy can reduce trading risk and maximise profits in the day-ahead market by adjusting its bidding strategy and the schedule for production or consumption. More specifically, a 1\% improvement in MAPE of forecast accuracy (within a 5\% to 14\% range) leads to about a 0.1 - 0.35\% cost reduction \cite{Zareipour2010254}. On average, a 1\% reduction in the MAPE of short-term price forecasts can result in savings of \$1.5 million per year for a typical medium-sized utility company with 5-GW peak load \cite{Uniejewski2016, LAGO2018890}. Furthermore, electricity is economically non-storable, and the imbalance between production and consumption can result in power system instability \cite{Vincent2013}. Accurate electricity forecasting allows energy firms to efficiently organise production or consumption, and this improves the stability of the power system.

In view of the findings from the analysis of feature impact, some implications are important for policy makers to improve cross-border trading in an integrated European power market: 
\begin{enumerate}
\item The external trading capacities from the German market play a salient role in the generation of Nordic electricity price, and an increasing influence\footnote{The NordLink, the power cable being built between Norway and Germany, is expected to commence operation in 2021.} is expected. Thus, all trading capacity between the Nordic and German markets allocated to Nord Pool for implicit auction in the day-ahead price formation could lead to a notable contribution to achieve better allocation of cross-border network capacity, such as the Nordic and Baltic bidding areas.

\item The German production prognosis has a significant predictive impact on the Nordic price. However, the progressive introduction of intermittent renewable energies in Germany\footnote{Renewable power generation covered more than 46\% of Germany's power consumption in 2020. The forecasting information of intermittent renewable energies, wind and photovoltaic infeed forecasts, in Germany are essential supply-side variables for the adjustment positions of intraday trading and are updated every 15 minutes until the physical delivery of electricity \citep{LI2021100189}.} makes it difficult to yield accurate predictions. Thus, it is essential to establish formal obligations for cross-border markets to collaborate by sharing useful prospective information. It can better serve the effective demand needed of the electricity trade.

\item The presence of physical transmission constraints can cause a substantial disruption to the market integration and the network congestion implies the shrinkage of the commercial capacity. Thus, an optimal network at a European level should be constructed by a decisive plan.
\end{enumerate}

\section{Conclusion} \label{S:9}
In this paper, we present three LSTM-based hybrid architectures for the EPF. This study puts emphasis on the influence of feature selection methods in the proposed hybrid models. In particular, we compare the prediction performance of the two-step feature selection, the autoencoder, and two-stage feature selection models based on the empirical study on the Nord Pool day-ahead system price. In addition, we employ a SHAP method to evaluate the importance and impact of the features on predicting this price. The main findings are the following: (1) We conclude that the different feature selection methods will lead to different feature selections. As input, diverse features will have a comparably significant impact on the performance of LSTM-based predictive models. (2) Compared to CNN-LSTM and ConvLSTM, LSTM-LSTM is a better autoencoder structure for EPF. (3) The two-stage models can improve the forecasting accuracy of two-step models to some extent. The superior feature selection from the RFE-SVR model allows the autoencoder model to detect more meaningful information for more accurate predictions. (4) The features from the German market (with the most power cables linking to Nord Pool) are more significant for EPF than others. This indicates that more interconnections will increase the cross-border influence on EPF. (5) Compared to other features, the exchange rates are relatively less important. (6) Flow deviation cannot significantly contribute to the price prediction because of its lack of variability. In many cases, the expected flow capacity is fully occupied. The network congestion implies that more interconnections are expected for an efficient Europe-wide electricity market.

For future studies, several extensions of the current study can be developed. Indeed, although the forecasting performance of the proposed models is considerable, we did not conduct an extensive grid search to optimise hyperparameters. It is reasonable to believe that the LSTM-based models with more comprehensive architecture will achieve better forecasting performance. The results will benefit spot electricity traders and policymakers, who make decisions based on accurate price predictions. Moreover, we envision that more testing on other feature selection models can obtain more and different feature selection subsets. They can provide more possibilities for researchers and industries to understand how different features affect prediction accuracy. Finally, the study was carried out using the data from the Nord Pool market, but the generality of the proposed models ensures a possible application to other integrated markets, such as EPEX and OMIE.

\section*{Acknowledgement}
This work acknowledges research support by COST Action “Fintech and Artificial Intelligence in Finance - Towards a transparent financial industry” (FinAI) CA19130, and has been performed within the +CityxChange\footnote{https://cityxchange.eu/} (Positive City ExChange) project under the Smart Cities and Communities topic that has received funding from the European Union’s Horizon 2020 research and innovation programme under Grant Agreement No. 824260. Critical comments and advice from Florentina Paraschiv, Rüdiger Kiesel,  and Frode Kjærland are gratefully acknowledged. The computations were performed on resources provided by UNINETT Sigma2 - the National Infrastructure for High Performance Computing and Data Storage in Norway.

\nomenclature[Z]{\(\text{EPF}\)}{Electricity price forecasting}
\nomenclature[Z]{\(\text{LSTM}\)}{Long short-term memory}
\nomenclature[Z]{\(\text{DNNs}\)}{Deep neural networks}
\nomenclature[Z]{\(\text{FNNs}\)}{Forward neural networks}
\nomenclature[Z]{\(\text{RNNs}\)}{Recurrent neural networks}
\nomenclature[Z]{\(\text{CNNs}\)}{Convolutional neural networks}
\nomenclature[Z]{\(\text{GRUs}\)}{Gated recurrent units}
\nomenclature[Z]{\(\text{PC}\)}{Pearson’s Correlation}
\nomenclature[Z]{\(\text{PSO}\)}{Particle swarm optimisation}
\nomenclature[Z]{\(\text{GA}\)}{Genetic algorithm}
\nomenclature[Z]{\(\text{ELM}\)}{Extreme learning machine}
\nomenclature[Z]{\(\text{ConvLSTM}\)}{Convolutional LSTM}

\nomenclature[Z]{\(\text{DM}\)}{Diebold-Mariano}
\nomenclature[Z]{\(\text{MAE}\)}{Mean absolute error}
\nomenclature[Z]{\(\text{RMSE}\)}{Root mean squared error}
\nomenclature[Z]{\(\text{MAPE}\)}{Mean absolute percentage error}
\nomenclature[Z]{\(\text{SMAPE}\)}{Symmetric mean absolute percentage error}
\nomenclature[Z]{\(\text{MSE}\)}{Mean squared error}

\nomenclature[Z]{\(\text{SHAP}\)}{SHapley Additive exPlanations}
\nomenclature[Z]{\(\text{NARMAX}\)}{Moving Average with eXogenous Input}



\newpage
\appendix
\section{}
\subsection{PSO}\label{PSO}
Each particle has knowledge about its current velocity, its own past best solution ($\overrightarrow{p}(t)$), and the current global best solution ($\overrightarrow{g}(t)$). Based on this information, each particle's velocity is updated such that it moves closer to the global best and its past best solution at the same time. The velocity update is performed according to the following equation:

\begin{equation}
\begin{split}
\overrightarrow{v}(t+1) & =\omega \overrightarrow{v}(t) + c_{1}r_1(\overrightarrow{p}(t)-\overrightarrow{x}(t))\\ &+  c_{2}r_2(\overrightarrow{g}(t)-\overrightarrow{x}(t)) 
\end{split}
\end{equation}
\noindent where $c_1$ and  $c_2$ are constants defined beforehand, which determine the significance of $\overrightarrow{p}(t)$ and $\overrightarrow{g}(t)$. $\overrightarrow{v}(t)$ is the velocity of the particle, $\overrightarrow{x}(t)$ is the current particle position, $r_1$ and $r_2$ are random numbers from the interval [0,1], and $\omega$ is a constant ($0 \leq \omega \le 1$).
The new position is calculated by summing the previous position and the new velocity as follows:
\begin{equation}
\overrightarrow{x}(t+1) = \overrightarrow{x}(t) +\overrightarrow{v}(t+1)
\end{equation}
This iterative process is repeated until a stopping criterion is satisfied.

\nomenclature[P]{\(\omega\)}{Constant that determines the significance of $\overrightarrow{v}$}
\nomenclature[P]{\(c_1\)}{Constant that determines the significance of $\overrightarrow{p}$}
\nomenclature[P]{\(c_2\)}{Constant that determines the significance of $\overrightarrow{g}$}
\nomenclature[P]{\(\overrightarrow{p}\)}{Particle's past best solution}
\nomenclature[P]{\(\overrightarrow{g}\)}{Global current best solution}
\nomenclature[P]{\(\overrightarrow{v}\)}{Velocity of the particle}
\nomenclature[P]{\(r_1, r_2\)}{Random number from the interval [0, 1]}

\subsection{GA-ELM workflow}\label{workflow}
\begin{figure}[t!]
  \centering
\includegraphics[width=0.5\textwidth]{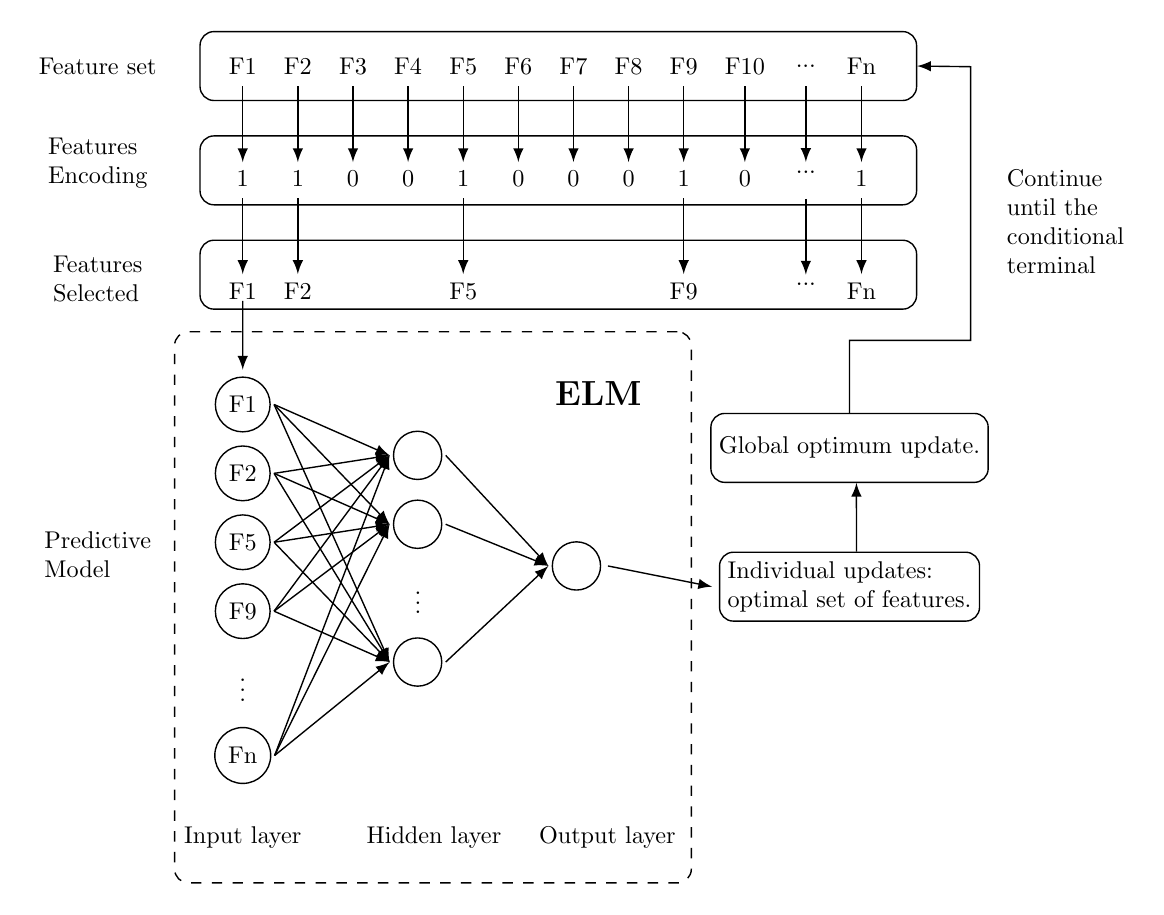}
\caption{The workflow of the two-step wrapper-based feature selection model.}\label{f9}
\end{figure}

Figure \ref{f9} shows the workflow of PSO-ELM and GA-ELM models for feature selection. The process flow of GA-ELM can be described as follows: 

\begin{steps}
  \item Initialise the population with a set of random individuals, each individual representing a particular subset of features. For a specific individual (feature set), the features are encoded as "1" or "0", as shown in Figure \ref{f9}. "1" means that the feature is selected, and  "0" means that it is not selected.
  \item The selected features are the input for the ELM. The prediction results of the ELM are used to evaluate the fitness value of the individuals. The fitness value is calculated based on the MSE.
  \item Select the best individual with regard to the fitness value. If its fitness is higher than the lowest value in the existing mating pool, it will replace the individual with the worst fitness. Furthermore, the global optimum will be updated accordingly.
  \item The child individuals are generated by crossover and mutation. The new generation is composed of a set of new individuals that are encoded and prepared to be evaluated. The whole process continues until meeting the iteration terminal. The best feature subset in the mating pool is the optimal selection.
\end{steps}

\subsection{SVR}\label{SVR}
\begin{figure}[t!]
\centering
\includegraphics[, height=5cm]{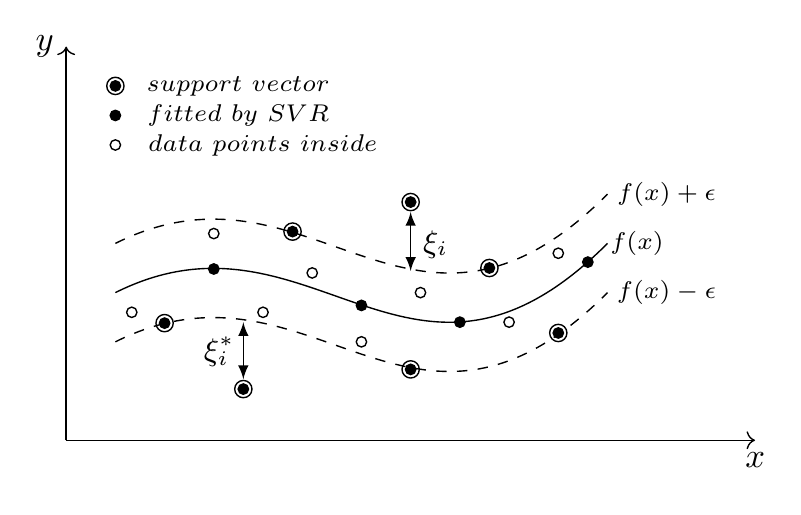}
\caption{\centering Fitted SVR.}\label{f10}
\end{figure}
To minimise the forecasting errors, SVR individualises the hyperplane by maximising the margin. To solve a nonlinear regression problem, the following linear estimation function is considered as follows \cite{HERCEG201995}:
\begin{equation}
f(x) = (\textbf{w}\times \Phi(x)) + b
\end{equation}
\noindent where $\textbf{w}$ is the parameter vector, $\Phi(x)$ is a kernel function and $b$ is a bias vector. The function formulation of the SVR model can be transformed into the following convex minimisation problem:
\begin{equation}
\min \frac{1}{2}||\textbf{w}||^2 + C\sum_{i=1}^{n}(\xi_i +\xi_i^*)
\end{equation}
\noindent subject to the following constraints:
\begin{equation*}
\begin{split}
(\textbf{w}\times \Phi(x_i) +b) -y_i \leq \varepsilon + \xi_i \\
y_i -(\textbf{w}\times \Phi(x_i) +b) \leq \varepsilon +\xi_i^* \\
\xi_i, \xi_i^* \geq 0; i = 1,2,...,n
\end{split}
\end{equation*}
where C is a regularisation constant and $\xi_i$ and $\xi_i^*$ are slack variables, which are used to handle the situation where no such function $f(x)$ exists to satisfy the constraint $|y_i -(\textbf{w}\times x_i +b)|\leq \varepsilon$ for all points. They are regarded as the soft margin to allow regression errors, $\varepsilon$, to exist up to $\xi_i$ and $\xi_i^*$ and still satisfy the constraint. Only the points outside the $\varepsilon$-radius contribute to the final cost. The error parameter, $\varepsilon$, represents the region of the tube located around the regression function, $f(x)$, as shown in Figure \ref{f10}.

\nomenclature[P]{\(\textbf{w}\)}{Parameter vector}
\nomenclature[P]{\(\Phi\)}{Kernel function}
\nomenclature[P]{\(\xi, \xi_i^*\)}{Slack variable}
\nomenclature[P]{\(C\)}{Regularisation constant}

\subsection{The statistical details of the model performance}
\begin{table}[h!]
\centering
\caption{\\The MAD (\%) results for M1, M2, M3, M4, M5, M6, M7 and M8.}\label{t11}
 \resizebox{0.5\textwidth}{!}{
 \begin{tabular}{c c c c c c c c c  } 
\toprule
 &M1 &M2&M3&M4&M5&M6&M7&M8\\
\midrule
count&10&10&10&10&10&10&10&10\\
mean&2.78&2.99&3.31&2.54&2.61&2.67&3.67&3.28\\
std&0.17&0.29&0.40&0.25&0.30&0.25&0.15&0.38\\
min&2.42&2.58&2.64&2.16&2.04&2.20&3.30&2.84\\
25\%&2.72&2.77&3.10&2.34&2.45&2.59&3.65&2.99\\
50\%&2.77&2.95&3.25&2.59&2.65&2.65&3.69&3.18\\
75\%&2.92&3.24&3.56&2.63&2.74&2.85&3.73&3.60\\
max&2.98&3.41&3.93&2.99&3.00&3.09&3.89&3.95\\
\bottomrule
\end{tabular}} 
\begin{tablenotes}
      \small
      \item Note:  25\%, 50\%, and 75\% denote 25\%, 50\%, and 75\% percentiles.
    \end{tablenotes}
\end{table}

\begin{table}[h!]
\centering
\caption{\\The RMSE (\%) results for M1, M2, M3, M4, M5, M6, M7 and M8.}\label{t12}
 \resizebox{0.5\textwidth}{!}{
 \begin{tabular}{c c c c c c c c c  } 
\toprule
 &M1 &M2&M3&M4&M5&M6&M7&M8\\
\midrule
count&10&10&10&10&10&10&10&10\\
mean&3.60&3.79&4.22&3.25&3.46&3.33&4.74&3.99\\
std&0.24&0.42&0.55&0.33&0.43&0.32&0.29&0.46\\
min&3.17&3.15&3.35&2.61&2.79&2.85&4.13&3.50\\
25\%&3.51&3.45&3.87&3.04&3.27&3.13&4.62&3.60\\
50\%&3.64&3.78&4.16&3.30&3.50&3.30&4.90&3.92\\
75\%&3.71&4.05&4.55&3.45&3.71&3.47&4.91&4.33\\
max&3.96&4.51&5.11&3.73&4.08&3.91&4.97&4.77\\
\bottomrule
\end{tabular}} 
\begin{tablenotes}
      \small
      \item Note:  25\%, 50\%, and 75\% denote 25\%, 50\%, and 75\% percentiles.
    \end{tablenotes}

\end{table}

\begin{table}[h!]
\centering
\caption{\\The MAPE (\%) results for M1, M2, M3, M4, M5, M6, M7 and M8.}\label{t13}
 \resizebox{0.5\textwidth}{!}{
 \begin{tabular}{c c c c c c c c c  } 
\toprule
 &M1 &M2&M3&M4&M5&M6&M7&M8\\
\midrule
count&10&10&10&10&10&10&10&10\\
mean&7.24&7.83&8.75&6.66&6.79&7.01&9.31&8.73\\
std&0.46&0.73&1.05&0.65&0.78&0.68&0.47&1.06\\
min&6.28&6.82&6.89&5.72&5.32&5.67&8.56&7.50\\
25\%&7.05&7.22&8.19&6.18&6.35&6.84&9.08&7.90\\
50\%&7.24&7.71&8.62&6.73&6.85&6.95&9.16&8.44\\
75\%&7.61&8.49&9.35&6.87&7.15&7.45&9.50&9.59\\
max&7.80&8.93&10.37&7.86&7.91&8.09&10.27&10.66\\
\bottomrule
\end{tabular}} 
\begin{tablenotes}
      \small
      \item Note:  25\%, 50\%, and 75\% denote 25\%, 50\%, and 75\% percentiles.
\end{tablenotes}

\end{table}

\begin{table}[h!]
\centering
\caption{\\The SMAPE (\%) results for M1, M2, M3, M4, M5, M6, M7 and M8.}\label{t_smape_table}
 \resizebox{0.5\textwidth}{!}{
 \begin{tabular}{c c c c c c c c c  } 
\toprule
 &M1 &M2&M3&M4&M5&M6&M7&M8\\
\midrule
count&10&10&10&10&10&10&10&10\\
mean&7.42&7.58&8.46&6.76&6.85&6.87&9.51&8.38\\
std&0.50&0.67&0.95&0.65&0.72&0.60&0.44&0.90\\
min&6.49&6.63&6.77&5.76&5.46&5.71&8.49&7.38\\
25\%&7.22&7.10&7.95&6.33&6.43&6.77&9.38&7.59\\
50\%&7.34&7.48&8.30&6.87&6.91&6.85&9.67&8.22\\
75\%&7.81&8.10&9.05&7.01&7.37&7.27&9.77&9.13\\
max&8.13&8.70&10.09&7.87&7.74&7.75&9.95&9.93\\
\bottomrule
\end{tabular}} 
\begin{tablenotes}
      \small
      \item Note:  25\%, 50\%, and 75\% denote 25\%, 50\%, and 75\% percentiles.
    \end{tablenotes}
\end{table}


\newpage
\bibliographystyle{model3-num-names} 
\bibliography{mybibfile}


%
%
%
\end{document}